\documentclass{article}

\PassOptionsToPackage{numbers}{natbib}

\usepackage[preprint]{neurips_2019}

\usepackage[utf8]{inputenc} 
\usepackage[T1]{fontenc}    
\usepackage{hyperref}       
\usepackage{url}            
\usepackage{booktabs}       
\usepackage{amsfonts}       
\usepackage{nicefrac}       
\usepackage{microtype}      

\usepackage{amsmath}
\usepackage{graphicx}
\usepackage{algorithm}
\usepackage{algorithmicx}

\usepackage{pgfplots}
\usepackage{pgfplotstable}
\usepgfplotslibrary{fillbetween}
\usepackage{wrapfig}

\usepackage{tikz}
\usetikzlibrary{decorations.pathreplacing}
\usepackage{caption}

\usepackage{algpseudocode}

\usepackage{mathtools}
\mathtoolsset{showonlyrefs}

\let\OldSim\sim
\renewcommand{\sim}{\,\normalsize{\OldSim}\,}

\newcommand{\sind}[1]{ {\scriptscriptstyle{#1}} }  

\newcommand{\T}{\mathsf{T}}

\title{Parameter elimination in particle Gibbs sampling}

\author{Anna Wigren \\
  Department of Information Technology\\
  Uppsala University, Sweden \\
  \texttt{anna.wigren@it.uu.se} \\
   \And
   Riccardo Sven Risuleo \\
   Department of Information Technology\\
   Uppsala University, Sweden \\
   \texttt{riccardo.risuleo@it.uu.se} \\
   \And
   Lawrence Murray \\
   Uber AI  \\
   San Francisco, CA, USA \\
   \texttt{lawrence.murray@uber.com} \\
   \And
   Fredrik Lindsten \\
   Division of Statistics and Machine Learning \\
   Linköping University, Sweden \\
   \texttt{fredrik.lindsten@liu.se} \\
}

\begin{document}

\maketitle

\definecolor{yellow1}{rgb}{1,0.55,0} 
\definecolor{green1}{rgb}{0,0.4,0}
\definecolor{blue1}{rgb}{0,0.1880,1}
\pgfplotsset{
	every axis/.append style={
		ylabel near ticks,
		xlabel near ticks,
		legend pos=north east,
		y tick label style={/pgf/number format/skip 0.},
		width=0.5\textwidth,
		height=0.34\textwidth,
		legend cell align=left,
		legend columns=-1,
		legend style={font=\footnotesize},
		cycle list name=exotic,
 		legend style={nodes={scale=0.85, transform shape}},
		legend image code/.code={\draw[mark repeat=3,mark phase=2] plot coordinates {(0cm,0cm) (0.15cm,0cm) (0.3cm,0cm)};} 
	}
}

\begin{abstract}
Bayesian inference in state-space models is challenging due to high-dimensional state trajectories. A viable approach is particle Markov chain Monte Carlo, combining MCMC and sequential Monte Carlo to form ``exact approximations'' to otherwise intractable MCMC methods. The performance of the approximation is limited to that of the exact method. We focus on particle Gibbs and particle Gibbs with ancestor sampling, improving their performance beyond that of the underlying Gibbs sampler (which they approximate) by marginalizing out one or more parameters. This is possible when the parameter prior is conjugate to the complete data likelihood. Marginalization yields a non-Markovian model for inference, but we show that, in contrast to the general case, this method still scales linearly in time. While marginalization can be cumbersome to implement, recent advances in probabilistic programming have enabled its automation. We demonstrate how the marginalized methods are viable as efficient inference backends in probabilistic programming, and demonstrate with examples in ecology and epidemiology.
\end{abstract}

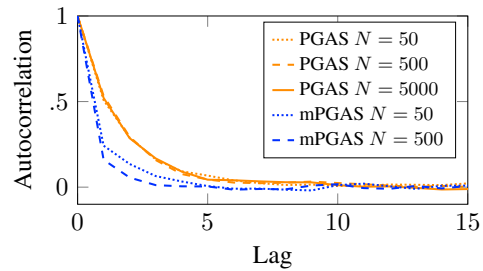
\begin{wrapfigure}{r}{6.1cm}
	\centering
	\vspace{-3ex} 
		\begin{tikzpicture}[scale=.96, transform shape]
  \pgfplotstableread[col sep=comma]{data/acfdata_new.csv}{\acfdata}
  \begin{axis}[
    xmin=0,
    xmax=15,
    ymin=-0.1,
    ymax=1,
    xlabel=Lag,
    ylabel=Autocorrelation,
    height=0.3\textwidth,
    legend columns=1,
    ]
    \addplot[thick, densely dotted, yellow1] table[x=lag, y=PGAS_50] {\acfdata};
    \addplot[thick, dashed, yellow1] table[x=lag, y=PGAS_500] {\acfdata};
    \addplot[thick, yellow1] table[x=lag, y=PGAS_5000] {\acfdata};
    \addplot[thick, densely dotted, blue1] table[x=lag, y=PGASm_50] {\acfdata};
    \addplot[thick, dashed, blue1] table[x=lag, y=PGASm_500] {\acfdata};
    \legend{PGAS $N=50$, PGAS $N=500$, PGAS $N=5000$, mPGAS $N=50$, mPGAS $N=500$};
  \end{axis}
\end{tikzpicture}
	\vspace{-3.1ex} 
	\caption{The autocorrelation function (ACF) for standard PGAS converges to that of the hypothetical Gibbs sampler as $N\rightarrow\infty$, whereas mPGAS will produce iid draws in the limit, i.e., the ACF will drop to zero at lag one for large $N$.
	Similar results hold for PG and mPG, see Supplementary \ref{sec:AddResult}.}
	\vspace{-3ex}
	\label{fig:acfPGAS}
\end{wrapfigure}

\vspace{-5mm}
\section{Introduction}
State-space models (SSMs) are a well-studied topic with applications in climatology \cite{CalafatWLWF:2018}, robotics \cite{DeisenrothNP:2013}, ecology \cite{Parslow2013}, and epidemiology \cite{RasmussenRK:2011}, to mention just a few. In this paper we propose a new method for performing Bayesian inference in such models. In SSMs, a latent (hidden) state process $x_\sind{t}$ is observed through a second process $y_\sind{t}$. The state process is assigned an initial density $x_\sind{0} \sim p(x_\sind{0})$, and evolves in time according to a transition density $x_\sind{t}\sim p(x_\sind{t}|x_\sind{t-1},\theta)$, where $\theta$ are parameters with prior density $p(\theta)$. Given the latent states $x_\sind{t}$, the observations are assumed independent with density $p(y_\sind{t}|x_\sind{t},\theta)$. We wish to infer the joint posterior, $p(x_\sind{0:T},\theta|y_\sind{1:T})$, of the states $x_\sind{0:T}$ and the parameters $\theta$, given a set of observations $y_\sind{1:T}=\{ y_\sind{1}, \dots , y_\sind{T} \}$. Unfortunately, computing this posterior distribution exactly is not analytically tractable for general non-linear, non-Gaussian models, so we must resort to approximations.

Markov chain Monte Carlo (MCMC) \cite[e.g.][]{RobertC:2004} is a popular choice for Bayesian inference. The motivation behind our new method is based on one such MCMC method: the Gibbs sampler. In the Gibbs sampler, samples from the posterior $p(x_\sind{0:T},\theta|y_\sind{1:T})$ are generated by alternating between sampling the states from $x_\sind{0:T}'\sim p(x_\sind{0:T}|y_\sind{1:T},\theta')$, and the parameters from $\theta' \sim p(\theta|x'_\sind{0:T},y_\sind{1:T})$. Sampling the parameters is often manageable, but sampling the states is challenging, owing to the distribution being high-dimensional. A possible remedy is to use particle Markov chain Monte Carlo (PMCMC) methods \cite{Andrieu10}, in which sequential Monte Carlo (SMC) is used to approximate sampling from the high-dimensional distribution. Particle Gibbs (PG) \cite{Andrieu10} is a PMCMC algorithm that mimics the Gibbs sampler. Efficient extensions, such as particle Gibbs with ancestor sampling (PGAS), have also been proposed, reducing the computational cost from quadratic to linear in the number of timesteps, $T$, in favorable conditions \cite{LeeSV:2018,Lindsten14}.

PG and PGAS have proven to be efficient in many challenging situations \cite[e.g.][]{LindermanSA:2014,MarcosCBD:2015,ValeraRSP:2015,MeentHMW:2015}. Nevertheless, being ``exact approximations''~\citep{Andrieu2009} of the (possibly intractable) Gibbs sampler, they can never outperform it. In essence, this means that when the number of particles used in their SMC component approaches infinity, PG and PGAS will approach the hypothetical Gibbs sampler in terms of autocorrelation, but can never surpass it. This is illustrated in Figure \ref{fig:acfPGAS}, orange curve (for details on the model, see Section \ref{sec:mPG}). Ideally, independent samples from the target distribution are desired, but the often strong dependence between the parameters $\theta$ and the states $x_\sind{0:T}$ in the hypothetical Gibbs sampler leads to correlated samples also for PG and PGAS.

In marginalized Gibbs sampling, we propose to marginalize out the parameters in the state update, ideally alternating between sampling the states $x_\sind{0:T}'\sim p(x_\sind{0:T}|y_\sind{1:T})$ and sampling the parameters $\theta'\sim p(\theta|x'_\sind{0:T},y_\sind{1:T})$ (note that an alternative is to sample only the state trajectories $\{x^\sind i_\sind{0:T}\}_\sind {i=1}^\sind M$, where $M$ is the number of MCMC steps, and then estimate the posterior of $\theta$ as a mixture of densities, where each component is $p(\theta | x^\sind i_\sind{0:T}, y_\sind{0:T})$). The state update is thus independent of the parameters and this hypothetical marginalized Gibbs sampler will effectively generate independent samples from the target distribution. However, like for the unmarginalized hypothetical Gibbs sampler, the distribution for sampling the states is not available in closed form. To address this issue, we derive marginalized versions of PG and PGAS (hereon referred to as mPG/mPGAS). Analogously to the unmarginalized case, when the number of particles go to infinity, mPG and mPGAS will approach the hypothetical marginalized Gibbs sampler -- that is, a sampler generating independent samples from the target. This behavior is illustrated in Figure \ref{fig:acfPGAS}, blue curve.

Marginalization is possible if the SSM has a conjugacy relation between the parameter prior and the complete data likelihood, that is, the conditional $p(\theta|x_\sind{0:T},y_\sind{1:T})$ has the same functional form as the prior $p(\theta)$. However, even for such models there is a price to pay for marginalization: it turns the Markovian dependencies, central to the SSM when conditioned on the parameters, into non-Markovian dependencies for both states and observations. This will make it harder to apply conventional MCMC methods, whereas PMCMC methods have proven to be better suited for models of this type \cite{Lindsten14}. In Section \ref{sec:method} we derive the algorithmic expressions for mPG and mPGAS for this family of models. The necessary updates in each step in the marginalized SMC algorithm can be done using sufficient statistics, which enables the computation time of mPG and mPGAS to scale linearly with the number of observations, despite the non-Markovian dependencies. The class of conjugate SSMs includes many common models, but is still somewhat restrictive. In Section \ref{sec:limitations}, we discuss some extensions to make the framework more generally applicable and provide numerical illustrations.

Marginalization of static parameters in the context of SMC has been studied by \cite{CarvalhoJLP:2010,Storvik:2002} for the purpose of online Bayesian parameter learning. To what extent these methods suffer from the well-known path degeneracy issue of SMC has been a topic of debate, see e.g. \cite{ChopinIMMRRS:2010}. Since our proposed method is based on PMCMC, and in particular PGAS, it is more robust to path degeneracy, see \cite{Lindsten14}. The Rao-Blackwellized particle filter \cite{ChenL:2000,DoucetFMR:2000} also makes use of marginalization, but for marginalizing part of the state vector using conditional Kalman filters.

In practice, deriving the conjugacy relations can be quite involved. However, recent developments in probabilistic programming have enabled automatic marginalization~\citep[see e.g.][]{Hoffman2018,Murray18,Obermeyer2019}, which significantly improves the usability of our proposed method. Probabilistic programming considers the way in which probabilistic models and inference algorithms may be expressed in universal programming languages, formally extending the expressive power of graphical models. There are by now quite a number of probabilistic programming languages. Examples that can support SMC-based methods, such as those considered here, include LibBi~\citep{Murray2015}, BiiPS~\citep{Todeschini2014}, Venture~\citep{Mansinghka2014}, Anglican~\citep{Tolpin2016}, WebPPL~\citep{Goodman2014},
Figaro~\citep{Pfeffer2016}, Turing~\citep{Ge2018}, and Birch~\citep{Murray2018a}. A language can implement PG/PGAS combined with automatic marginalization to realize our proposed method. We have implemented PG, mPG, PGAS and mPGAS in Birch \cite{Murray2018a} and provide examples to illustrate their efficiency in Section \ref{sec:PPL} and \ref{sec:mPart}.

\section{Background on SMC} \label{sec:SMC}
In PG and PGAS, the state update is approximated using SMC, therefore we provide a brief summary of the SMC algorithm before introducing the proposed method. For a more extensive introduction, see e.g. \cite{Cappe07,Gordon93}. Consider a sequence of probability densities $\bar{\gamma}_\sind{\theta,t}(x_\sind{0:t})$ expressed as
\begin{equation} \label{eq:SMCtarget}
	\bar{\gamma}_\sind{\theta,t}(x_\sind{0:t})=\frac{\gamma_\sind{\theta,t}(x_\sind{0:t})}{Z_\sind{\theta,t}}, \hspace{5mm} t=1,2,\dots
\end{equation}
where $\gamma_\sind{\theta,t}$ are the corresponding unnormalized densities, which we assume can be evaluated pointwise, and $Z_\sind{\theta,t}$ is a normalizing constant. For a SSM, the target density of interest is often $p(x_\sind{0:t}|y_\sind{1:t},\theta)$, which implies $\gamma_\sind{\theta,t}=p(x_\sind{0:t},y_\sind{1:t}|\theta)$ and $Z_\sind{\theta,t}= p(y_\sind{1:t}|\theta)$.
SMC methods approximate the target density \eqref{eq:SMCtarget} using a set of $N$ weighted samples (or particles) $\{ x_\sind{0:t}^\sind{i}, \bar{w}_\sind{t}^\sind{i} \}_{i=1}^N$,
generated according to Algorithm \ref{alg:SMC}.
  When moving to the next distribution in the sequence, all particles are resampled by choosing an ancestor trajectory $x_\sind{0:t-1}^\sind{a_t^i}$ from the previous step in time according to the respective weights $\bar{w}_\sind{t-1}^i$ of the possible ancestors. SMC is based on importance sampling and the resampled particles are therefore propagated to the next time step using a proposal distribution, $q_\sind{\theta,t}(x_\sind{t}|x_\sind{0:t-1})$, chosen by the user. A common choice for SSMs is to use the
  bootstrap proposal, which equates to propagating according to the transition density $p(x_\sind{t}|x_\sind{t-1},\theta)$, but other more refined choices, such as the optimal proposal (see e.g. \cite{Doucet00}), are also possible. Finally, the (unnormalized) importance weights for the propagated particles are computed using the weight function 
\begin{equation} \label{eq:wSMC}
	\omega_\sind{\theta,t}(x_\sind{0:t}) = \frac{\gamma_\sind{\theta,t}(x_\sind{0:t})}{\gamma_\sind{\theta,t-1}(x_\sind{0:t-1})q_\sind{\theta,t}(x_\sind{t}|x_\sind{0:t-1})}.
\end{equation}

\begin{algorithm}[H]
	\caption{SMC (all steps for $i=1,\dots,N$)}
	\label{alg:SMC}
	\begin{algorithmic}[1] 
		\State \textit{Initialize:} Draw $x_\sind{0}^\sind{i} \sim q_\sind{0}(x_\sind{0})$, set $w_\sind{0}^i=\gamma_\sind{\theta,0}(x_\sind{0}^\sind{i})/q_\sind{0}(x_\sind{0}^\sind{i})$, normalize $\bar{w}_\sind0^\sind i=w_\sind0^\sind i/\sum_{j=1}^N w_\sind0^\sind j$
		\For {$t=1 \dots T$}
		\State \textit{Resample:} Draw $a_\sind t^\sind i\sim \mathcal{C}(\{\bar{w}_\sind{t-1}^\sind i\}_{i=1}^N)$, where $\mathcal{C}$ is the categorical distribution.
		\State \textit{Propagate:} Simulate $x_\sind t^\sind i \sim q_\sind{\theta,t}(x_\sind t|x_\sind{0:t-1}^\sind{a_t^i})$.
		\State \textit{Update:} Set $w_\sind t^\sind i=\omega_\sind{\theta,t}(x_\sind{0:t}^\sind i)$ according to \eqref{eq:wSMC} and normalize $\bar{w}_\sind t^\sind i=w_\sind t^\sind i/\sum_{j=1}^N w_\sind t^\sind j$
		\EndFor
	\end{algorithmic}
\end{algorithm}

\section{Method} \label{sec:method}
In this section, we first specify the class of models we consider, and then we show how to marginalize the SMC algorithm and derive mPG and mPGAS for this class of models.

\subsection{Conjugate models and marginalized SMC} \label{sec:mSMC}
The SMC framework presented in Section \ref{sec:SMC} is in a general form and can be directly applied to the marginalized state update by defining the unnormalized target distribution as $\gamma_\sind{t}(x_\sind{0:t})=p(x_\sind{0:t},y_\sind{1:t})$ in \eqref{eq:SMCtarget} and then applying Algorithm \ref{alg:SMC}. The computation of the importance weights (step 5 in Algorithm~\ref{alg:SMC}), however, turns out to be problematic in marginalized SSMs. To see why, note that the unnormalized target density can be factorized into $p(x_\sind{0:t},y_\sind{1:t})=p(x_\sind0)\prod_{k=1}^{t}p(x_\sind k,y_\sind k|x_\sind{0:k-1},y_\sind{1:k-1})$. The weights \eqref{eq:wSMC} become
\begin{equation} \label{eq:margW}
	\omega_\sind{t}(x_\sind{0:t}) = \frac{p(x_\sind t,y_\sind t|x_\sind{0:t-1},y_\sind{1:t-1})}{q_\sind t(x_\sind t|x_\sind{0:t-1})}
\end{equation}
where the numerator (and possibly also the denominator depending on the choice of proposal) is non-Markovian. The marginal joint density of states and observations can be written
\begin{equation} \label{eq:mGenInt}
	p(x_\sind t, y_\sind t|x_\sind{0:t-1},y_\sind{1:t-1}) = \int p(x_\sind t,y_\sind t|x_\sind{t-1},\theta)p(\theta|x_\sind{0:t-1},y_\sind{1:t-1})\mathrm{d}\theta
\end{equation}
where $p(\theta|x_\sind{0:t-1},y_\sind{1:t-1})$ is the posterior distribution of the parameters.
For a general SSM, the integral \eqref{eq:mGenInt} is intractable, and the posterior may be difficult to compute. However, if there is a conjugacy relationship between the prior distribution $p(\theta)$ and the complete data likelihoods $p(x_\sind{0:t},y_\sind{1:t}|\theta), \hspace{1mm} t=1,\dots,T$, the integral can be solved analytically and the posterior will be of the same form as the prior. One such case is when both the complete data likelihood and the parameter prior are in the \emph{exponential family}, see Supplementary~\ref{sec:Expfam} for details. However, if we consider joint state and observation likelihoods, $p(x_\sind t,y_\sind t|x_\sind{t-1},\theta)$, in the exponential family, we can end up with a log-partition function that depends on the previous state $x_\sind{t-1}$. This can create problems when formulating a conjugate prior for the complete data likelihoods since the prior will be different for each state update, see Supplementary \ref{sec:RestExpfam} for details. To avoid this problem for the models we consider, we introduce the \emph{restricted exponential family} where the joint state and observation likelihood is given by
\begin{equation} \label{eq:modelExp}
p(x_\sind t,y_\sind t|x_\sind{t-1},\theta) = h_\sind t \exp\left( \theta^\T s_\sind t - A^\T(\theta)r_\sind t \right)
\end{equation}
where $h_\sind t=h(x_\sind{t},x_\sind{t-1},y_\sind{t})$ is the data dependent base measure, $s_\sind t=s(x_\sind{t},x_\sind{t-1},y_\sind{t})$ is a sufficient statistic and where the log-partition function can be separated into two factors: $A(\theta)$, which is independent of the data, and $r_\sind t=r(x_\sind{t-1})$, which is independent of the parameters.
A conjugate prior for \eqref{eq:modelExp} is given by
\begin{equation} \label{eq:priorExp}
	\pi(\theta|\chi_\sind0,\nu_\sind0) = g(\chi_\sind0,\nu_\sind0) \exp \left( \theta^\T \chi_\sind0 - A^\T(\theta)\nu_\sind0 \right)
\end{equation}	
where $\chi_\sind0$, $\nu_\sind 0$ are hyperparameters. The parameter posterior is given by $p(\theta|x_\sind{0:t-1},y_\sind{1:t-1}) = \pi(\theta|\chi_\sind{t-1},\nu_\sind{t-1})$, with the hyperparameters iteratively updated according to
\begin{equation} \label{eq:paramUpdate}
	\chi_\sind{t} = \chi_\sind0 + \sum_{k=1}^{t}s_\sind k = \chi_\sind{t-1}+s_\sind{t}, \hspace{10mm}  \nu_\sind{t} = \nu_\sind0 + \sum_{k=1}^{t}r_\sind k = \nu_\sind{t-1}+r_\sind{t}.
\end{equation}
With the joint likelihood \eqref{eq:modelExp} and its conjugate prior \eqref{eq:priorExp} 
in place, we can derive an analytic expression for the marginal of the joint distribution of states and observations, \eqref{eq:mGenInt}, at time $t$  
\begin{equation} \label{eq:margJoint}
	\begin{split}
		&p(x_\sind t,y_\sind t|x_\sind{0:t-1},y_\sind{1:t-1}) = \int p(x_\sind t,y_\sind t|x_\sind{t-1}\theta) \pi(\theta|\chi_\sind{t-1},\nu_\sind{t-1}) \mathrm{d}\theta = \frac{g(\chi_\sind{t-1},\nu_\sind{t-1})}{g(\chi_\sind t,\nu_\sind t)}h_\sind t.
	\end{split}
\end{equation}
Hence, to compute the weights \eqref{eq:margW} for marginalized SMC in the restricted exponential family, we only need to keep track of and update the hyperparameters according to \eqref{eq:paramUpdate}.

\subsection{Marginalized particle Gibbs}\label{sec:mPG}
In PG, we alternate between sampling the parameters and the states like in the hypothetical Gibbs sampler, but the state trajectory is sampled using conditional SMC (cSMC). In cSMC one particle trajectory, the reference trajectory $x_\sind{0:T}'$, will always survive the resampling step. This version of SMC follows the steps in Algorithm \ref{alg:SMC}, with the constraints that $a_\sind t^\sind{N}=N$ and  $x_\sind t^\sind{N}\,\normalsize{=}\,x_\sind t'$ (for details, see~\cite{Andrieu10}). When marginalizing out the parameters, the resulting mPG sampler updates the state trajectory using marginalized cSMC (mcSMC), according to what is presented in Algorithm \ref{alg:SMC} and Section \ref{sec:mSMC}, with the addition of conditioning on the reference trajectory surviving the resampling step (like in standard PG). 

The conditioning used in cSMC yields a Markov kernel that leaves the correct conditional distribution invariant for any choice of $N$ \cite{Andrieu10}. PG is therefore a valid MCMC procedure. However, it has been shown that $N$ must increase (at least) linearly with $T$ for the kernel to mix properly for large $T$, resulting in an overall computational complexity which grows quadratically with $T$. This holds also for other popular PMCMC methods, such as particle marginal Metropolis-Hastings \cite{Andrieu10}. To mitigate this issue, \cite{Lindsten14} proposed a modification of PG in which the ancestor for the reference trajectory in each time step is sampled, according to ancestor weights $\tilde{w}^\sind i_\sind{t-1|T}$, instead of set deterministically, which significantly improves the mixing of the kernel for small $N$, even when $T$ is large. The resulting method, referred to as PGAS, is equivalent to PG apart from the resampling step.

The difference between mPG and mPGAS lies, analogous to the non-marginalized case, only in the resampling step. Deriving the expression for the ancestor weights in the marginalized case is quite involved, below we simply state the necessary expressions and updates, a complete derivation is provided in Supplementary \ref{sec:ASderiv}. Each ancestor trajectory in mPGAS is assigned a weight, based on the general expression in \cite{Lindsten14}, given by
\begin{equation} \label{eq:wPGASm}
\tilde{w}^\sind i_\sind{t-1|T}=\bar{w}^\sind i_\sind{t-1} 
\frac{\gamma_\sind{T}([x^\sind i_\sind{0:t-1},x'_\sind{t:T}])}{\gamma_\sind{t-1}(x^\sind i_\sind{0:t-1})}
= \bar{w}^\sind i_\sind{t-1} \frac{p([x^\sind i_\sind{0:t-1},x'_\sind{t:T}],y_\sind{1:T})}{p(x^\sind i_\sind{0:t-1},y_\sind{1:t-1})},
\end{equation}
where $\bar{w}^\sind i_\sind{t-1}$ is the weight of the ancestor trajectory $x^\sind i_\sind{0:t-1}$ and $[x^\sind i_\sind{0:t-1},x'_\sind{t:T}]$ is the concatenated trajectory resulting from combining the reference trajectory $x'_\sind{t:T}$ with the possible ancestral path $x^\sind i_\sind{0:t-1}$. For members of the restricted exponential family we use \eqref{eq:margJoint} in \eqref{eq:wPGASm} to get the weights
\begin{equation}\label{eq:wPGASexp}
\tilde{w}^\sind i_\sind{t-1|T} \propto \bar{w}^\sind i_\sind{t-1}h_\sind t^\sind i \frac{g(\chi_\sind{t-1} ^\sind i,\nu_\sind{t-1} ^\sind i )}{g(\chi_\sind t ^\sind i,\nu_\sind t^\sind i)} \prod_{k=t+1}^{T}h'_\sind k \frac{g(\chi_\sind{k-1} ^\sind i,\nu_\sind{k-1} ^\sind i )}{g(\chi_\sind k ^\sind i,\nu_\sind k^\sind i)} \propto \bar{w}^\sind i_\sind{t-1} \frac{g(\chi_\sind{t-1}^\sind i,\nu_\sind{t-1}^\sind i)}{g(\chi_\sind{T}^\sind i,\nu_\sind{T}^\sind i)}h_\sind{t}^\sind i,
\end{equation}
where $\chi_\sind{t-1}^\sind i$, $\nu_\sind{t-1}^\sind i$ are given, for each particle, by \eqref{eq:paramUpdate} and where
\begin{align} \label{eq:paramAS}
 \chi_\sind{T}^\sind i=\chi_\sind{t-1}^\sind i + s_\sind{t}(x'_\sind{t},x^\sind i_\sind{t-1},y_\sind{t}) + s'_\sind{t+1:T} \, , \qquad
 \nu_\sind{T}^\sind i=\nu_\sind{t-1}^\sind i + r_\sind{t}(x^\sind i_\sind{t-1}) + r'_\sind{t+1:T}\, ,
\end{align}
with $s'_\sind{t+1:T}=\sum_{k=t+1}^{T}s_\sind{k}(x'_\sind{k},x'_\sind{k-1},y_\sind{k})$ and similarly for $r_\sind{t}$. Hence, $\chi_\sind{T}^\sind i$ is a combination of the statistic for the ancestor trajectory, a cross-over term and the statistic for the reference trajectory, which in each timestep is updated according to $s'_\sind{t+1:T}=s'_\sind{t:T}-s_\sind{t}(x'_\sind{t},x'_\sind{t-1},y_\sind{t})$, and analogously for $\nu_\sind{T}^\sind i$ and $r'_\sind{t+1:T}$. By storing and updating these parameters and sum of statistics in each iteration, computing the ancestor sampling weights only amounts to evaluating \eqref{eq:wPGASexp}, implying that we can run mPGAS in linear time despite having a non-Markovian target, which would normally yield quadratic complexity (see \cite{Lindsten14} for a discussion). We outline mPGAS in Algorithm \ref{alg:PGAS} (for mPG, skip step 3, updates of $\chi_\sind T$, $\nu_\sind T$ and set $a^N_\sind t$ deterministically).
\begin{algorithm}[H]
	\caption{Marginalized PGAS for the restricted exponential family (all steps for $i=1,\dots,N$)}
	\label{alg:PGAS}
	\begin{algorithmic}[1]
		\Require $x_\sind{0:T}'$, $s_\sind{1:T}'$, $r_\sind{1:T}'$
		\State \textit{Initialize:} Draw $x_\sind{0}^\sind{1:N-1} \sim q_\sind{0}(x_\sind{0})$, set $x_\sind{0}^\sind{N}=x'_\sind{0}$, set $w_\sind{0}^\sind{i}=\frac{\gamma_{0}(x_{0}^{i})}{q_0(x_0^i)}$ and $\bar{w}_\sind{0}^\sind{i}=\frac{w_0^i}{\sum_{j=1}^N w_0^j}$
		\For {$t=1 \dots T$}
		\State \textit{Update statistics:} $s'_\sind{t+1:T}=s'_\sind{t:T}-s_\sind{t}(x'_\sind{t},x'_\sind{t-1},y_\sind{t})$,  $r'_\sind{t+1:T}=r'_\sind{t:T}-r_\sind{t}(x'_\sind{t-1})$
		\State \textit{Update hyperparameters:} $\chi_\sind{t}^\sind{i}$, $\nu_\sind{t}^\sind{i}$, $\chi^\sind i_\sind{T}$, $\nu^\sind i_\sind{T}$ according to \eqref{eq:paramUpdate} and \eqref{eq:paramAS} 
		\State \textit{Resample:} Draw $a_\sind{t}^\sind{1:N-1} \sim \mathcal{C}(\{\bar{w}_\sind{t-1}^\sind{i}\}_{i=1}^N)$ and $a_\sind{t}^\sind{N} \sim \mathcal{C}(\{\tilde{w}_\sind{t-1|T}^\sind{i} \}_{i=1}^N)$, $\tilde{w}_\sind{t-1|T}^\sind{i}$ from \eqref{eq:wPGASexp}
		\State \textit{Propagate:} Simulate $x_\sind{t}^\sind{1:N-1} \sim q_\sind{t}(x_\sind{t}|x_\sind{0:t-1}^\sind{a_t^{1:N-1}})$ and set $x_\sind{t}^\sind{N}=x_\sind{t}'$
		\State \textit{Update weights:} Set $w_\sind{t}^\sind{i}=\omega_\sind{t}(x_\sind{0:t}^\sind{i})$ according to \eqref{eq:margW} and normalize $\bar{w}_\sind{t}^\sind{i}=w_\sind{t}^\sind{i}/\sum_{j=1}^N w_\sind{t}^\sind{j}$
		\EndFor
		\Ensure Sample new $x_\sind{0:T}'$, $s_\sind{1:T}'$, $r_\sind{1:T}'$ according to $\bar{w}_\sind T $
	\end{algorithmic}
\end{algorithm}

To illustrate the improved performance offered by marginalization we consider the non-linear SSM \cite{Gordon93}
\begin{align} \label{eq:toyModel}
x_\sind{t} = \frac{x_\sind{t-1}}{2} + 25\frac{x_\sind{t-1}}{1+x_\sind{t-1}^{2}} + 8\cos (1.2t) + v_\sind{t}, \hspace{10mm} y_\sind{t} = \frac{x_\sind{t}^{2}}{20} + w_\sind{t},
\end{align}
where $v_\sind{t}$ and $w_\sind{t}$ are Gaussian noise processes with zero mean and unknown variances $\sigma_\sind{v}^{2}$ and $\sigma_\sind{w}^{2}$ respectively. The observations are a quadratic function of the state, which makes the posterior multimodal. We will assume conjugate, inverse gamma priors $\sigma_\sind{v}^{2} \sim \mathcal{IG}(\alpha_\sind v, \beta_\sind v)$ and $\sigma_\sind{w}^{2} \sim \mathcal{IG}(\alpha_\sind w, \beta_\sind w)$ for the unknown variances, with hyperparameters $\alpha_\sind v = \beta_\sind v = \alpha_\sind w = \beta_\sind w =1$. We generated $T=150$ observations from \eqref{eq:toyModel} with $\sigma_\sind{v}^{2}=10$ and $\sigma_\sind{w}^{2}=1$.
PGAS and mPGAS were run for $M=10000$ iterations, discarding the first 1500 samples as burn-in.
We initialized with $\sigma_\sind v^{2}=\sigma_\sind w^2=100$ and used a bootstrap proposal for PGAS and a marginalized bootstrap proposal for mPGAS.

Figure \ref{fig:acfPGAS} shows the autocorrelation for PGAS and mPGAS for different number of particles~$N$. Ideally we would like iid samples from the posterior distribution, in terms of the ACF of the samples it should be zero everywhere except for lag 0. It is clear that, for PGAS, increasing $N$ can reduce the autocorrelation only to a certain limit (given by the hypothetical Gibbs sampler). For mPGAS on the other hand, we obtain a lower autocorrelation using only 50 particles as compared to 5000 for PGAS, and by increasing $N$ we move towards generating iid samples. In Supplementary \ref{sec:AddResult} we provide corresponding results for PG/mPG. The results in Figure \ref{fig:acfPGAS} were obtained from our implementation in Matlab, in Supplementary \ref{sec:AddResult} we also show the corresponding results for our implementation in Birch.

The marginalized versions of PG/PGAS requires some extra computations compared to their non-marginalized counterparts, however, this overhead is quite small. For the model \eqref{eq:toyModel}, with N=500, using the tic-toc timer in MATLAB we get: PG -- 1231.5s, mPG -- 1430.7s, PGAS -- 1260.7s, mPGAS -- 1566.1s. Note that the code has not been optimized.
\section{Extensions and numerical simulations} \label{sec:limitations}
In this section we describe three extensions of the marginalized method presented in Section \ref{sec:method} and illustrate their efficiency in numerical examples.

\subsection{Diffuse priors and blocking} \label{sec:block}
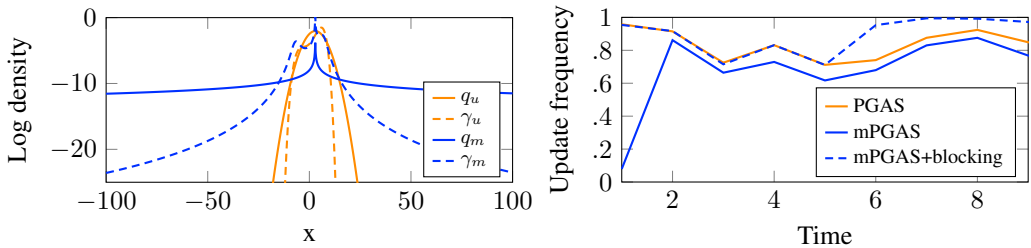
\begin{figure}[b]
	\begin{center}
		\begin{tikzpicture}[scale=1, transform shape]
  \pgfplotstableread[col sep=comma]{data/PDFdata.csv}{\PDFdata}
  \pgfplotstableread[col sep=comma]{data/UpdateFreqdata.csv}{\acfdata}
  \begin{axis}[
    xmin=-100,
    xmax=100,
    ymin=-25,
    ymax=0,
    xlabel=x,
    ylabel=Log density,
    width=0.5\textwidth,
   	height=0.27\textwidth,
    legend columns=1,
    legend pos=south east,
    name=myplot,
    ]
    \addplot+[mark=none,thick,yellow1] table[x=x1, y=log_q_um] {\PDFdata};
    \addplot+[mark=none,thick,densely dashed, yellow1] table[x=x1, y=log_post_um] {\PDFdata};
    \addplot+[mark=none,thick, blue1] table[x=x1, y=log_q_m] {\PDFdata};
    \addplot+[mark=none,thick,densely dashed, blue1] table[x=x1, y=log_post_m] {\PDFdata};
    \legend{$q_u$, $\gamma_u$ , $q_m$, $\gamma_m$};
  \end{axis}
  \begin{axis}[
	  xmin=1,
 	 xmax=9,
 	 ymin=0,
 	 ymax=1,
 	 xlabel=Time,
 	 ylabel=Update frequency,
 	 width=0.5\textwidth,
 	 height=0.27\textwidth,
 	 legend columns=1,
 	 legend pos=south east,
 	 at=(myplot.outer east),
 	 anchor=outer west,
 	 ]
	 \addplot+[mark=none,thick, color=yellow1] table[x=time, y=PGAS] {\acfdata};
	 \addplot+[mark=none,thick, color=blue1] table[x=time, y=PGASm] {\acfdata};
	 \addplot+[mark=none,thick,densely dashed, color=blue1] table[x=time, y=PGASmb] {\acfdata};
	 \legend{PGAS, mPGAS, mPGAS+blocking};
  \end{axis}
\end{tikzpicture}
	\end{center}
	\caption{\emph{Left:} log-density for the proposal and the posterior at $t=1$ for mcSMC ($q_\sind m$, $\gamma_\sind m$) and cSMC ($q_\sind u$, $\gamma_\sind u$), showing how marginalization can potentially produce a poor proposal distribution in the first timestep. \emph{Right:} update frequency for the state trajectory for the first few timesteps.}\label{fig:blockplot}
\end{figure}

When we do not know much about the parameters of a model, we may use a diffuse prior to reflect our uncertainty. However, a diffuse prior on the parameters can lead to a diffuse prior also for the states. We can then encounter problems during the first few timesteps of the marginalized state trajectory update; in particular, if we use a bootstrap-style proposal in the mcSMC algorithm it may spread out the particles too much. This can result in poor mixing during the initial timesteps, as well as numerical difficulties in the computation of the ancestor sampling weights, due to very large values sampled for the states. As an illustration, consider again the model \eqref{eq:toyModel}, but now with hyperparameters $\alpha_\sind v,\beta_\sind v = 0.001$ for the process noise $\sigma_\sind v ^2$. The marginalized proposal for the first timestep, $q_\sind1(x_\sind1|x_\sind0)=p(x_\sind 1\mid x_\sind0)$, will then be a Student $t$-distribution with undefined mean and variance. Figure \ref{fig:blockplot} (left) shows the log-pdf of both this proposal and the target distribution, $\bar{\gamma}_\sind{1}(x_\sind{0:1})=p(x_\sind{0:1} \mid y_\sind 1)$, at time $t=1$. It is clear that for mcSMC (blue) the prior $q_\sind 1$ is much more diffuse than the posterior $\bar{\gamma}_\sind{1}$, whereas for cSMC (orange) there is less of a difference.

When working with diffuse priors we suggest to divide the state trajectory into two overlapping blocks (similarly to the blocking method proposed by \cite{SinghLM:2017}) and do Gibbs updates of each block in turn. Figure \ref{fig:Blocking} illustrates the two overlapping blocks $x_\sind{0:B+L}$ (upper) and $x_\sind{B+1:T}$ (lower). To update the first block, where problems due to marginalization are more probable, we use a (non-marginalized) cSMC sampler targeting the posterior distribution of $x_\sind{0:B+L}$ conditioned on the reference trajectory $x'_\sind{0:B+L}$, the observations $y_\sind{1:T}$, the non-overalapping part of the second block $x'_\sind{B+L+1:T}$ and the parameters $\theta$. Note that, because of the Markov property when conditioning on $\theta$, the dependence on $x_\sind{0:B+L}$ reduces to only the boundary state $x'_\sind{B+L+1}$ and the dependence on the observations reduces to $y_\sind{1:B+L}$. To update the second block, we use mcSMC targeting the posterior distribution of $x_\sind{B+1:T}$ conditioned on the (updated) reference trajectory $ [ x_\sind{B+1:B+L},x'_\sind{B+L+1:T} ]$, the observations $y_\sind{1:T}$ and the (updated) first block $ x_\sind{0:B}$. Finally, the parameters $\theta$ are sampled from their full conditional given the new reference trajectory $x_\sind{0:T}$. Algorithm \ref{alg:Blocking} outlines one iteration of mPG/mPGAS with this choice of blocking and samplers. In Supplementary \ref{sec:JustBlocking} we provide a proof of validity for this blocked Gibbs sampler.

The purpose of the first block is only to update the first few timesteps, in order to get a sufficient concentration of the proposals when conditioning on $x_\sind{0:B}$ for mcSMC. Therefore, it is typically sufficient to use a small value of $B$; in the example outlined above $B>2$ is sufficient to get finite variance in the Student $t$-distribution. The overlap parameter $L$ on the other hand is used to push the boundary state $x_\sind{B+L+1}$ into the interior of the second block, which due to the forgetting of the dynamical system reduces the effect of conditioning on this state in the first Gibbs step \cite{SinghLM:2017}. Hence, the larger $L$ the better, but at the price of increased computational cost. Since most SSMs have exponential forgetting, using a small value of $L$ is likely to be sufficient in most cases.

In Figure \ref{fig:blockplot} (right), we illustrate the benefit of using blocking to avoid poor mixing during the first timestep when marginalizing with a diffuse prior for the model \eqref{eq:toyModel}. We used $B=5$ and $L=20$, all other settings were the same as before.
We consider the update frequency of the state variables, defined as the average number of iterations in which the state changes its value, as a measure of the mixing. It is clear that for the mPGAS we get a very low update frequency at $t=1$, whereas when we use mPGAS with blocking we obtain the same update frequency as for PGAS.

\noindent\begin{minipage}{0.62\textwidth}
	\begin{algorithm}[H]
		\caption{Blocking for mPG/mPGAS}
		\label{alg:Blocking}
		\begin{algorithmic}[1] 
			\State $x_\sind{0:B+L} \sim \text{cSMC}(x_\sind{0:B+L}|x'_\sind{0:B+L};y_\sind{1:B+L},x'_\sind{B+L+1},\theta)$
			\State $x_\sind{B+1:T} \sim \text{mcSMC}(x_\sind{B+1:T}|x_\sind{B+1:B+L},x'_\sind{B+L+1:T};x_\sind{0:B},y_\sind{1:T})$
			\State $\theta \sim p(\theta|x_\sind{0:T},y_\sind{1:T})=\pi(\theta|\chi_\sind{T},\nu_\sind{T})$
		\end{algorithmic}
	\end{algorithm}

	\vspace{3mm}
\end{minipage}
\begin{minipage}[c]{0.38\textwidth}
	\hspace{1mm}
	\resizebox{0.98\textwidth}{!}{
		\begin{tikzpicture}[scale=1, transform shape, ]
			\def\B{1}
			\def\BL{7}
			\def\T{10}
			\draw[thin, black, fill=yellow1!50] (0,0) node[above] at ++ (0,0.75) {$0$} rectangle (\B,0.75)
			node[above] { $B$};
			\draw[thin, black, fill=yellow1!50] (\B,0) rectangle
			(\BL,0.75) node[above] { $B+L$ };
			\draw[thin, black, dashed] (\BL,0) rectangle (\T,0.75) node[above] {$T$};
			\draw[decorate, decoration={brace, amplitude=8pt}, yshift=-.6em] (\B,0) to
			node[midway,yshift=-1.6em] { \Large $x_\sind{0:B}$ } (0,0);
			\draw[decorate, decoration={brace, amplitude=8pt}] (\BL,0) to
			node[midway,yshift=-1.6em] {\Large $x_\sind{0:B+L}$ } (0,0);

			\draw[thin, black, dashed] (0,-1.85) rectangle (\B,-1.1);
			\draw[thin, black, fill=yellow1!50] (\B,-1.85) rectangle (\T,-1.1);
			\draw[thin, black, fill=yellow1!50] (\B,-1.85) rectangle (\BL,-1.1);

			\draw[decorate, decoration={brace, amplitude=8pt}, yshift=-.6em] (\T,-1.85) to
			node[midway,yshift=-1.6em] { \Large $x_\sind{B+L+1:T}$} (\BL,-1.85);
			\draw[decorate, decoration={brace, amplitude=8pt}] (\T,-1.85) to
			node[midway,yshift=-1.6em] {\Large $x_\sind{B+1:T}$} (\B,-1.85);
		\end{tikzpicture}
	}
	
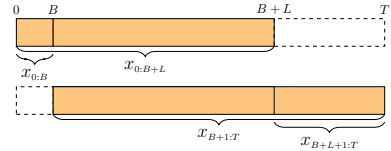
\captionof{figure}{Division into 2 blocks.}
	\label{fig:Blocking}
\end{minipage}

\vspace{-3mm}

\subsection{Marginalized particle Gibbs in a PPL} \label{sec:PPL}
We have implemented PG, PGAS, mPG and mPGAS in Birch \citep{Murray2018a}, which employs \emph{delayed sampling}~\citep{Murray18} to recognize and utilize conjugacy relationships, and so automatically marginalizes out the parameters of a model, where possible. This saves the user the trouble of deriving the relevant conjugacy relationships for their particular model, or providing a bespoke implementation of them. We first demonstrate this on a vector-borne disease model of a dengue outbreak.

Dengue is a mosquito-borne disease which affects an estimated 50-100 million people worldwide each year, causing 10000 deaths~\citep{Stanaway2016}. We use a data set from an outbreak on the island of Yap in Micronesia in 2011. It contains 197 observations, mostly daily, of the number of newly reported cases. The model used is that described in \citep{Murray18}, in turn based on that of the original study \citep{Funk16}. It consists of two coupled susceptible-exposed-infectious-recovered (SEIR) compartmental models, describing the transmission between human and mosquito populations, respectively. Transition counts between compartments are assumed to be binomially distributed, with beta priors used for all parameters. Observations are also assumed binomial with an unknown parameter for the reporting rate, which is assigned a beta prior. The beta priors establish conjugate relations with the complete data likelihood, so that the problem is well-suited for inference using mPG/mPGAS.

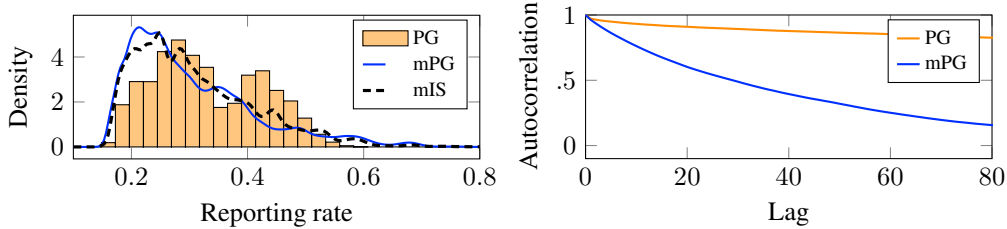
\begin{figure}[b]
	\begin{center}
		\begin{tikzpicture}[scale=1, transform shape]
  \pgfplotstableread[col sep=comma]{data/density.csv}{\datavbddensity}
  \pgfplotstableread[col sep=comma]{data/histogram.csv}{\datavbdhist}
  \pgfplotstableread[col sep=comma]{data/autocorr.csv}{\datavbdautocorr}
  \begin{axis}[
    xmin=0.1,
    xmax=0.8,
    xlabel=Reporting rate,
    width = 0.5\textwidth,
    height = 0.25\textwidth,
    ylabel=Density,
    legend columns = 1,
    legend pos = north east,
    name = myplot
    ]
    \addplot[ybar interval, fill=yellow1!50, area legend] table[x=rho, y=density_pg] {\datavbdhist};
    \addplot[thick, blue1] table[x=rho,y=density_mpg] {\datavbddensity};
    \addplot[very thick, densely dashed, black] table[x=rho, y=density_imp]{\datavbddensity};
    \legend{PG, mPG, mIS};
  \end{axis}
  \begin{axis}[
  	xmin=0,
  	xmax=80,
  	ymin=-0.1,
  	ymax=1,
  	xlabel=Lag,
  	width = 0.5\textwidth,
  	height = 0.25\textwidth,
  	ylabel=Autocorrelation,
	legend columns = 1,
	legend pos = north east,
	at = (myplot.north east),
	anchor=north west,
	xshift=4em,
  ]
  \addplot[thick, yellow1] table[x=lag, y=pg] {\datavbdautocorr};
  \addplot[thick, blue1] table[x=lag, y=mpg] {\datavbdautocorr};
  \legend{PG, mPG};
  \end{axis}
\end{tikzpicture}
	\end{center}
	\caption{Results of the simulation of the vector-borne disease model. \emph{Left:} estimated density of the reporting rate parameter, mean of four chains. Marginalized importance sampling (mIS) is included for comparison. Note that the marginalized methods yield mixture distributions.
		 \emph{Right:} estimated autocorrelation function of the reporting rate parameter, mean of four chains.}
	\label{fig:VBD}
\end{figure}

The model was previously implemented in Birch for \citep{Murray18}. We have added generic implementations of PG, PGAS, mPG and mPGAS to Birch that can be applied to this, and other, models. Figure \ref{fig:VBD} shows the results of a simulation of four different chains; for each of these 10000 samples were drawn using PG and mPG. The samplers used $N=1024$ particles each. For comparison we also include the results from using marginalized importance sampling. The autocorrelation of the samples is noticeably improved by marginalizing out the parameters. Corresponding results for PGAS and mPGAS can be found in Supplementary \ref{sec:AddResult}.

\subsection{Models lacking full conjugacy} \label{sec:mPart}
It may seem that the method we propose is limited to models where the transition and observation probabilities have the conjugacy structure in \eqref{eq:modelExp}. However, we can use the results in Section \ref{sec:method} to treat models where only some of the parameters exhibit conjugacy with the complete data likelihood. To this end, we denote by $\theta_\sind m$ the parameters that have a prior distribution that is conjugate with the complete-data likelihood, and by $\theta_\sind u$ the remaining parameters. Then, we can marginalize out $\theta_\sind m$ from the complete-data likelihood as shown in Section \ref{sec:method}.
The remaining parameters can be sampled using any conventional MCMC method, for instance Metropolis--Hastings. This is possible since PMCMC
samplers are nothing but (special purpose) MCMC kernels, hence they can be combined with normal MCMC in a systematic way. One possibility is to use, say, Metropolis--Hastings within mPG/mPGAS. Another possibility, which we describe below, is to use a marginalized version of the particle marginal Metropolis--Hastings algorithm \cite{Andrieu10}, which we refer to as mPMMH.

Let $\hat p(y_\sind{1:T}|\theta_\sind u)= \prod_{t=1}^{T}\frac{1}{N}\sum_{i=1}^{N} w_\sind t^\sind i$ be the unbiased estimate of the marginal likelihood given by Algorithm \ref{alg:SMC}, for a fixed value of the parameters $\theta_u$, and let $q(\theta_u |\theta_\sind u')$ be a proposal distribution; then, we can generate samples from the posterior distribution of $\theta_\sind u$ using Algorithm \ref{alg:PMMH}.
\begin{algorithm}[H]
	\caption{Marginalized particle marginal Metropolis--Hastings}
	\label{alg:PMMH}
	\begin{algorithmic}[1] 
		\State Propose $\theta_\sind u^\ast \sim q(\,\cdot\, | \theta_\sind u')$
		\State Run Algorithm \ref{alg:SMC} and compute $\hat p(y_\sind{1:T}|\theta_\sind u^\ast)$
		\State Return $\theta_\sind u=\theta^\ast_\sind u$ with probability 	$1\wedge \frac{\hat p( y_\sind{1:T}|\theta_\sind u^\ast)p(\theta_\sind u^\ast)q(\theta_\sind u'|\theta_\sind u^\ast)}{\hat p( y_\sind{1:T}|\theta_\sind u')p(\theta_\sind u')q(\theta_\sind u^\ast|\theta_\sind u')}$, else $\theta_\sind u = \theta_\sind u'$,
	\end{algorithmic}
\end{algorithm}
To illustrate this method with partial marginalization, we consider the following model describing the evolution of the size of animal populations (see, for instance, \cite{Lande03}):
\begin{equation}
\log n_\sind{t+1} = \log n_\sind t + \begin{bmatrix}
1 & {(n_\sind t)}^c
\end{bmatrix} b + \sigma_\sind v v_\sind t, \qquad y_\sind t =
n_\sind t + \sigma_\sind w w_\sind t,
\end{equation}
where $n_\sind t$ is the population size at time $t$, and $b$, $c$, $\sigma_\sind v$, and $\sigma_\sind w$ are the unknown parameters. Note that, except for $c$ ($=\theta_\sind u$), the parameters can be marginalized out by using normal-inverse gamma and inverse gamma conjugate priors $b,\sigma_\sind v^2 \sim \mathcal{NIG}(\mu,\Lambda,\alpha_\sind v,\beta_\sind v)$ and $\sigma_\sind{w}^2 \sim \mathcal{IG}(\alpha_\sind w, \beta_\sind w)$. For the remaining parameter, we use a $\mathcal{N}(0,\sigma_\sind c^2)$ prior and a random-walk proposal $c^\ast \sim \mathcal{N}(c',\tau)$.

\begin{figure}[H]
	\begin{center}
		\begin{tikzpicture}[scale=1, transform shape]
	\pgfplotstableread[col sep=comma] {data/ecological-histogram.csv}{\dataecohist};
	\pgfplotstableread[col sep=comma] {data/ecological-realization.csv}{\dataecoreal};
	\begin{axis}[xmin=-5,
		xmax=5,
		xlabel=$c$,
		ylabel=Density,
		width=0.4\textwidth,
   		height=0.25\textwidth,
		name=myplot]
		\addplot[ybar interval, fill=yellow1!50] table[x=b3, y=Pb3] {\dataecohist};
	\end{axis}
	\begin{axis}[
		xtick={1975, 1980, 1985,1990,1995},
		x tick label style={/pgf/number format/1000 sep={}},
		xmin=1974.5,
		xmax=1997.5,
    	height=0.25\textwidth,
		width=0.6\textwidth,
		xlabel=Year,
		ylabel=Population,
		at=(myplot.outer east),
		anchor=outer west,
		legend columns=1,
		legend pos=south east
		]
		\addplot+[mark=*, only marks,blue1] table[x=t, y=y] {\dataecoreal};
		\addplot+[thick, mark=none,yellow1] table[x=t, y=y_hat] {\dataecoreal};
		\addplot[gray!20, name path=upper] table[x=t, y expr=\thisrow{y_hat} + 3.0*\thisrow{sigma_y}] {\dataecoreal};
		\addplot[gray!20, name path=lower] table[x=t, y expr=\thisrow{y_hat} - 3.0*\thisrow{sigma_y}] {\dataecoreal};
		\addplot[gray!20] fill between[of=upper and lower];
		\legend{$n_t$, $\hat n_t$}
	\end{axis}
\end{tikzpicture}
	\end{center}
	\caption{Results of the simulation with parameter values $\mu=[1,\,1]$, $\Lambda=I$,
		$\alpha_\sind v=\beta_\sind v=\alpha_\sind w=\beta_\sind w=2.5$, $\sigma_\sind c^2 = 4$, $\tau=0.05$.
		\emph{Left:} estimated distribution of the density regulation
		parameter $c$. \emph{Right:} observed (marks) and mean filtered population
		sizes (solid) with $3\sigma$ credible interval.
	}\label{fig:sparrows}
\end{figure}
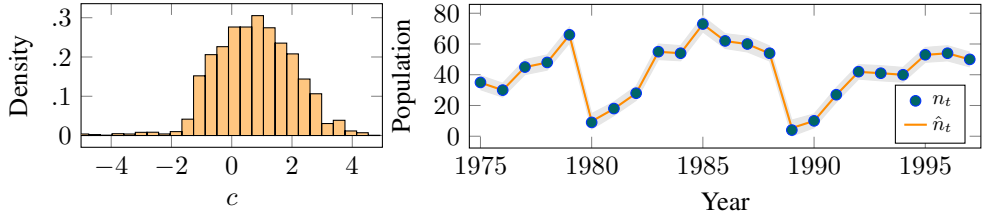

We have implemented mPMMH in Birch and evaluate it on a dataset of observations of the number of song sparrows on Mandarte Island, British Columbia, Canada \cite{Saether00}. The dataset contains the number of birds, counted yearly, between 1978 and 1998. In Figure \ref{fig:sparrows} (left), we report the histogram of the distribution of the density regulation parameter $c$ estimated using 10000 samples drawn using Algorithm \ref{alg:PMMH} after a burn-in of 5000 samples, using $N=512$ particles. The distribution of $c$, as found by our method, is consistent with values reported in the literature (see, for instance, \cite{Nadeem12} and \cite{Saether00}). In Figure~\ref{fig:sparrows} (right), we show the actual counts in the dataset compared with the average $\hat n_\sind {1:T}$ and three standard deviations, as sampled by Algorithm \ref{alg:PMMH}.

\section{Discussion}
PG and PGAS can be highly efficient samplers for general SSMs, but are limited by the performance of the hypothetical (but intractable) Gibbs sampler that they approximate. We have proposed to improve on PG/PGAS by marginalizing out the parameters from the state update, to reduce the auto-correlation beyond the limit posed by the hypothetical Gibbs sampler.

Marginalization often improves performance, but this will not always be the case. One example is when there is a diffuse prior on the parameters, in which case marginalization can result in an inefficient SMC sampler. One way to mitigate this is blocking; we propose using two blocks, the first updated using cSMC and the second using mcSMC. One can think of other ways to update the first block, such as a Metropolis--Hastings update with an appropriate proposal, see \cite{FearnheadM:2016,Murray2013a} for related techniques. It is also possible to use a mcSMC update for the first block, as conditioning on the future states will help to avoid the problems related to diffuse priors. The details are quite involved, however, so we prefer the simpler method described in Section \ref{sec:block}. 

Marginalization is possible when there is a conjugacy relationship between the parameters and the complete data likelihood. This may seem a restrictive model class, but in practice there are benefits even if only some of the parameters can be marginalized out, by combining marginalized PMCMC kernels with conventional MCMC kernels. Many models have at least some parameters that enter in a nice way, such as regression coefficients and error variances, where marginalization can provide a performance gain.

Performing the marginalization by hand for every new model can be time consuming. Consequently, an important aspect of the method is the possibility of implementing it in a probabilistic programming language. Recent advances in probabilistic programming enable automatic marginalization, making the process easier. We have implemented mPG, mPGAS and mPMMH in Birch, and demonstrated that implementation on two examples. Some further work is required to extend the implementation in Birch to blocking.

\subsubsection*{Code}
Code for all numerical simulations is available at \\ \url{https://github.com/uu-sml/neurips2019-parameter-elimination}.
\subsubsection*{Acknowledgments}
This research is financially supported, in part, by the Swedish Research Council via the projects \emph{Learning of Large-Scale Probabilistic Dynamical Models} (contract number: 2016-04278) and \emph{ New Directions in Learning Dynamical Systems (NewLEADS)} (contract number: 2016-06079), by the Swedish Foundation for Strategic Research (SSF) via the projects \emph{Probabilistic Modeling and Inference for Machine Learning} (contract number: ICA16-0015) and \emph{ASSEMBLE} (contract number: RIT15-0012), and by the Wallenberg AI, Autonomous Systems and Software Program (WASP) funded by the Knut and Alice Wallenberg Foundation.


\newpage

\bibliographystyle{abbrvnat}
\bibliography{refs}

\begin{thebibliography}{41}
\providecommand{\natexlab}[1]{#1}
\providecommand{\url}[1]{\texttt{#1}}
\expandafter\ifx\csname urlstyle\endcsname\relax
  \providecommand{\doi}[1]{doi: #1}\else
  \providecommand{\doi}{doi: \begingroup \urlstyle{rm}\Url}\fi

\bibitem[Andrieu and Roberts(2009)]{Andrieu2009}
C.~Andrieu and G.~O. Roberts.
\newblock The pseudo-marginal approach for efficient {M}onte {C}arlo
  computations.
\newblock \emph{Annals of Statistics}, 37\penalty0 (2):\penalty0 697--725,
  2009.

\bibitem[Andrieu et~al.(2010)Andrieu, Doucet, and Holenstein]{Andrieu10}
C.~Andrieu, A.~Doucet, and R.~Holenstein.
\newblock Particle {M}arkov chain {M}onte {C}arlo methods.
\newblock \emph{Journal of the Royal Statistical Society. Series B (Statistical
  Methodology)}, 72\penalty0 (3):\penalty0 269--342, 2010.

\bibitem[Calafat et~al.(2018)Calafat, Wahl, Lindsten, Williams, and
  Frajka-Williams]{CalafatWLWF:2018}
F.~M. Calafat, T.~Wahl, F.~Lindsten, J.~Williams, and E.~Frajka-Williams.
\newblock Coherent modulation of the sea-level annual cycle in the {U}nited
  {S}tates by {A}tlantic {R}ossby waves.
\newblock \emph{Nature Communications}, 9\penalty0 (2571), 2018.

\bibitem[Capp\'{e} et~al.(2007)Capp\'{e}, Godsill, and Moulines]{Cappe07}
O.~Capp\'{e}, S.~J. Godsill, and E.~Moulines.
\newblock An overview of existing methods and recent advances in sequential
  {M}onte {C}arlo.
\newblock \emph{Proceedings of the IEEE}, 95\penalty0 (5):\penalty0 899--924,
  2007.

\bibitem[Carvalho et~al.(2010)Carvalho, Johannes, Lopes, and
  Polson]{CarvalhoJLP:2010}
C.~M. Carvalho, M.~S. Johannes, H.~F. Lopes, and N.~G. Polson.
\newblock Particle learning and smoothing.
\newblock \emph{Statistical Science}, 25\penalty0 (1):\penalty0 88--106, 2010.

\bibitem[Chen and Liu(2000)]{ChenL:2000}
R.~Chen and J.~S. Liu.
\newblock Mixture {K}alman filters.
\newblock \emph{Journal of the Royal Statistical Society: Series B (Statistical
  Methodology)}, 62\penalty0 (3):\penalty0 493--508, 2000.

\bibitem[Chopin et~al.(2010)Chopin, Iacobucci, Marin, Mengersen, Robert, Ryder,
  and Schäfer]{ChopinIMMRRS:2010}
N.~Chopin, A.~Iacobucci, J.-M. Marin, K.~Mengersen, C.~P. Robert, R.~Ryder, and
  C.~Schäfer.
\newblock On particle learning.
\newblock arXiv.org, arXiv:1006.0554, 2010.

\bibitem[Deisenroth et~al.(2013)Deisenroth, Neumann, and
  Peters]{DeisenrothNP:2013}
M.~P. Deisenroth, G.~Neumann, and J.~Peters.
\newblock A survey on policy search for robotics.
\newblock \emph{Foundations and Trends in Robotics}, 2\penalty0
  (1--2):\penalty0 1--142, 2013.

\bibitem[Doucet et~al.(2000{\natexlab{a}})Doucet, De~Freitas, Murphy, and
  Russell]{DoucetFMR:2000}
A.~Doucet, N.~De~Freitas, K.~Murphy, and S.~Russell.
\newblock Rao-{B}lackwellised particle filtering for dynamic {B}ayesian
  networks.
\newblock In \emph{Proceedings of the 16th conference on Uncertainty in
  artificial intelligence}, pages 176--183, 2000{\natexlab{a}}.

\bibitem[Doucet et~al.(2000{\natexlab{b}})Doucet, Godsill, and
  Andrieu]{Doucet00}
A.~Doucet, S.~Godsill, and C.~Andrieu.
\newblock On sequential {M}onte {C}arlo sampling methods for {B}ayesian
  filtering.
\newblock \emph{Statistics and Computing}, 10\penalty0 (3):\penalty0 197--208,
  2000{\natexlab{b}}.

\bibitem[Fearnhead and Meligkotsidou(2016)]{FearnheadM:2016}
P.~Fearnhead and L.~Meligkotsidou.
\newblock Augmentation schemes for particle {MCMC}.
\newblock \emph{Statistics and Computing}, 26\penalty0 (6):\penalty0
  1293--1306, 2016.

\bibitem[Funk et~al.(2016)Funk, Kucharski, Camacho, Eggo, Yakob, Murray, and
  Edmunds]{Funk16}
S.~Funk, A.~J. Kucharski, A.~Camacho, R.~M. Eggo, L.~Yakob, L.~M. Murray, and
  W.~J. Edmunds.
\newblock Comparative analysis of dengue and {Z}ika outbreaks reveals
  differences by setting and virus.
\newblock \emph{PLOS Neglected Tropical Diseases}, 10\penalty0 (12):\penalty0
  1--16, 2016.

\bibitem[Ge et~al.(2018)Ge, Xu, and Ghahramani]{Ge2018}
H.~Ge, K.~Xu, and Z.~Ghahramani.
\newblock Turing: a language for flexible probabilistic inference.
\newblock In \emph{Proceedings of Machine Learning Research, Twenty-First
  International Conference on Artificial Intelligence and Statistics},
  volume~84, pages 1682--1690, 2018.

\bibitem[Goodman and Stuhlm\"{u}ller(2014)]{Goodman2014}
N.~D. Goodman and A.~Stuhlm\"{u}ller.
\newblock The design and implementation of probabilistic programming languages.
\newblock http://dippl.org, 2014.

\bibitem[Gordon et~al.(1993)Gordon, Salmond, and Smith]{Gordon93}
N.~J. Gordon, D.~J. Salmond, and A.~F.~M. Smith.
\newblock Novel approach to nonlinear/non-{G}aussian {B}ayesian state
  estimation.
\newblock In \emph{IEE proceedings F (radar and signal processing)}, volume
  140, pages 107--113, 1993.

\bibitem[Hoffman(2018)]{Hoffman2018}
M.~D. Hoffman.
\newblock Autoconj: recognizing and exploiting conjugacy without a
  domain-specific language.
\newblock In \emph{Advances in Neural Information Processing Systems (NeurIPS)
  31}, pages 10716--10726. 2018.

\bibitem[Lande et~al.(2003)Lande, Engen, and Saether]{Lande03}
R.~Lande, S.~Engen, and B.-E. Saether.
\newblock \emph{Stochastic population dynamics in ecology and conservation}.
\newblock Oxford University Press, Oxford, 2003.

\bibitem[Lee et~al.(2018)Lee, Singh, and Vihola]{LeeSV:2018}
A.~Lee, S.~S. Singh, and M.~Vihola.
\newblock Coupled conditional backward sampling particle filter.
\newblock arXiv.org, arXiv:1806.05852, 2018.

\bibitem[Linderman et~al.(2014)Linderman, Stock, and Adams]{LindermanSA:2014}
S.~Linderman, C.~H. Stock, and R.~P. Adams.
\newblock A framework for studying synaptic plasticity with neural spike train
  data.
\newblock In \emph{Advances in Neural Information Processing Systems ({NIPS})
  27}. 2014.

\bibitem[Lindsten et~al.(2014)Lindsten, Jordan, and Sch{\"o}n]{Lindsten14}
F.~Lindsten, M.~I. Jordan, and T.~B. Sch{\"o}n.
\newblock Particle {G}ibbs with ancestor sampling.
\newblock \emph{Journal of Machine Learning Research}, 15:\penalty0 2145--2184,
  2014.

\bibitem[Mansinghka et~al.(2014)Mansinghka, Selsam, and Perov]{Mansinghka2014}
V.~K. Mansinghka, D.~Selsam, and Y.~N. Perov.
\newblock Venture: a higher-order probabilistic programming platform with
  programmable inference.
\newblock arXiv.org, arXiv:1404.0099, 2014.

\bibitem[Marcos et~al.(2015)Marcos, Calafat, Berihuete, and
  Dangendorf]{MarcosCBD:2015}
M.~Marcos, F.~M. Calafat, A.~Berihuete, and S.~Dangendorf.
\newblock Long-term variations in global sea level extremes.
\newblock \emph{Journal of Geophysical Research}, 120\penalty0 (12):\penalty0
  8115--8134, 2015.

\bibitem[Murray(2015)]{Murray2015}
L.~M. Murray.
\newblock Bayesian state-space modelling on high-performance hardware using
  {LibBi}.
\newblock \emph{Journal of Statistical Software}, 67\penalty0 (10):\penalty0
  1--36, 2015.

\bibitem[Murray and Schön(2018)]{Murray2018a}
L.~M. Murray and T.~B. Schön.
\newblock Automated learning with a probabilistic programming language:
  {B}irch.
\newblock \emph{Annual Reviews in Control}, 46:\penalty0 29--43, 2018.

\bibitem[Murray et~al.(2013)Murray, Jones, and Parslow]{Murray2013a}
L.~M. Murray, E.~M. Jones, and J.~Parslow.
\newblock On disturbance state-space models and the particle marginal
  {M}etropolis--{H}astings sampler.
\newblock \emph{SIAM/ASA Journal of Uncertainty Quantification}, 1\penalty0
  (1):\penalty0 494--521, 2013.

\bibitem[Murray et~al.(2018)Murray, Lundén, Kudlicka, Broman, and
  Schön]{Murray18}
L.~M. Murray, D.~Lundén, J.~Kudlicka, D.~Broman, and T.~B. Schön.
\newblock Delayed sampling and automatic {R}ao-{B}lackwellization of
  probabilistic programs.
\newblock \emph{Proceedings of Machine Learning Research, Twenty-First
  International Conference on Artificial Intelligence and Statistics},
  84:\penalty0 1037--1046, 2018.

\bibitem[Nadeem and Lele(2012)]{Nadeem12}
K.~Nadeem and S.~R. Lele.
\newblock Likelihood based population viability analysis in the presence of
  observation error.
\newblock \emph{Oikos}, 121\penalty0 (10):\penalty0 1656--1664, 2012.

\bibitem[Obermeyer et~al.(2019)Obermeyer, Bingham, Jankowiak, Pradhan, Chiu,
  Rush, and Goodman]{Obermeyer2019}
F.~Obermeyer, E.~Bingham, M.~Jankowiak, N.~Pradhan, J.~Chiu, A.~Rush, and
  N.~Goodman.
\newblock Tensor variable elimination for plated factor graphs.
\newblock \emph{36th International Conference on Machine Learning (ICML)},
  2019.

\bibitem[Parslow et~al.(2013)Parslow, Cressie, Campbell, Jones, and
  Murray]{Parslow2013}
J.~Parslow, N.~Cressie, E.~P. Campbell, E.~Jones, and L.~M. Murray.
\newblock Bayesian learning and predictability in a stochastic nonlinear
  dynamical model.
\newblock \emph{Ecological Applications}, 23\penalty0 (4):\penalty0 679--698,
  2013.

\bibitem[Pfeffer(2016)]{Pfeffer2016}
A.~Pfeffer.
\newblock \emph{Practical probabilistic programming}.
\newblock Manning, 2016.

\bibitem[Rasmussen et~al.(2011)Rasmussen, Ratmann, and
  Koelle]{RasmussenRK:2011}
D.~A. Rasmussen, O.~Ratmann, and K.~Koelle.
\newblock Inference for nonlinear epidemiological models using genealogies and
  time series.
\newblock \emph{PLoS Comput Biology}, 7\penalty0 (8), 2011.

\bibitem[Robert and Casella(2004)]{RobertC:2004}
C.~P. Robert and G.~Casella.
\newblock \emph{{M}onte {C}arlo statistical methods}.
\newblock Springer, 2004.

\bibitem[Saether et~al.(2000)Saether, Engen, Lande, Arcese, and
  Smith]{Saether00}
B.-E. Saether, S.~Engen, R.~Lande, P.~Arcese, and J.~N.~M. Smith.
\newblock Estimating the time to extinction in an island population of song
  sparrows.
\newblock \emph{Proceedings: Biological Sciences}, 267\penalty0
  (1443):\penalty0 621--626, 2000.

\bibitem[Singh et~al.(2017)Singh, Lindsten, and Moulines]{SinghLM:2017}
S.~S. Singh, F.~Lindsten, and E.~Moulines.
\newblock Blocking strategies and stability of particle {G}ibbs samplers.
\newblock \emph{Biometrika}, 104\penalty0 (4):\penalty0 953--969, 2017.

\bibitem[Stanaway et~al.(2016)Stanaway, Shepard, Undurraga, Halasa, Coffeng,
  Brady, Hay, Bedi, Bensenor, Casta\~{n}eda Orjuela, Chuang, Gibney, Memish,
  Rafay, Ukwaja, Yonemoto, and Murray]{Stanaway2016}
J.~D. Stanaway, D.~S. Shepard, E.~A. Undurraga, Y.~A. Halasa, L.~E. Coffeng,
  O.~J. Brady, S.~I. Hay, N.~Bedi, I.~M. Bensenor, C.~A. Casta\~{n}eda Orjuela,
  T.-W. Chuang, K.~B. Gibney, Z.~A. Memish, A.~Rafay, K.~N. Ukwaja,
  N.~Yonemoto, and C.~J.~L. Murray.
\newblock The global burden of dengue: an analysis from the {Global Burden of
  Disease Study} 2013.
\newblock \emph{The Lancet Infectious Diseases}, 16\penalty0 (6):\penalty0
  712--723, 2016.

\bibitem[Storvik(2002)]{Storvik:2002}
G.~Storvik.
\newblock Particle filters for state-space models with the presence of unknown
  static parameters.
\newblock \emph{IEEE Transactions on Signal Processing}, 50\penalty0
  (2):\penalty0 281--289, 2002.

\bibitem[Todeschini et~al.(2014)Todeschini, Caron, Fuentes, Legrand, and
  Del~Moral]{Todeschini2014}
A.~Todeschini, F.~Caron, M.~Fuentes, P.~Legrand, and P.~Del~Moral.
\newblock Biips: software for {B}ayesian inference with interacting particle
  systems.
\newblock arXiv.org, arXiv:1412.3779, 2014.

\bibitem[Tolpin et~al.(2016)Tolpin, van~de Meent, Yang, and Wood]{Tolpin2016}
D.~Tolpin, J.~van~de Meent, H.~Yang, and F.~Wood.
\newblock Design and implementation of probabilistic programming language
  {A}nglican.
\newblock arXiv.org, arXiv:1608.05263, 2016.

\bibitem[Valera et~al.(2015)Valera, Ruiz, Svensson, and
  Perez-Cruz]{ValeraRSP:2015}
I.~Valera, F.~Ruiz, L.~Svensson, and F.~Perez-Cruz.
\newblock Infinite factorial dynamical model.
\newblock In \emph{Advances in Neural Information Processing Systems ({NIPS})
  28}. 2015.

\bibitem[van~de Meent et~al.(2015)van~de Meent, Hongseok, Mansinghka, and
  Wood]{MeentHMW:2015}
J.-W. van~de Meent, Y.~Hongseok, V.~Mansinghka, and F.~Wood.
\newblock Particle {G}ibbs with ancestor sampling for probabilistic programs.
\newblock In \emph{Proceedings of the 18th International Conference on
  Artificial Intelligence and Statistics}, 2015.

\bibitem[van Dyk and Park(2008)]{vanDyk08}
D.~A. van Dyk and T.~Park.
\newblock Partially collapsed {G}ibbs samplers.
\newblock \emph{Journal of the American Statistical Association}, 103\penalty0
  (482):\penalty0 790--796, 2008.

\end{thebibliography}
\newpage 

\section*{Supplementary material}

\appendix

\section{\normalsize Exponential family} \label{sec:Expfam}
A generic exponential family distribution can be written
\begin{equation} \label{eq:ExpLhood}
	p(z|\eta) = h(z) \exp\left( \eta^\T s(z) - a(\eta)\right)
\end{equation}
where $h$ is the data dependent base measure, $a$ is the log-partition function, and $s$ is a sufficient statistic storing all the information about the natural parameters $\eta$ contained in the data $z$. The conjugate prior for an exponential family distribution is also in the exponential family and is given by
\begin{equation} \label{eq:ExpPrior}
	\pi(\eta|\chi,\nu)=g(\chi,\nu)\exp\left(\eta^\T\chi-a(\eta)\nu\right)
\end{equation}
where $g$ is a normalizing factor and  $\chi$, $\nu$ are hyperparameters. The parameter posterior for a prior \eqref{eq:ExpPrior} and a likelihood \eqref{eq:ExpLhood} is given by
\begin{align} \label{eq:ExpPosterior}
	p(\eta|z,\chi,\nu)\propto p(z|\eta)p(\eta|\chi,\nu) = h(z)g(\chi,\nu)\exp\left( \eta^\T \left(\chi+s(z)\right)-a(\eta)(\nu+1) \right)
\end{align}
where Bayes' rule was used in the first proportionality. If we compare the exponential factor in the posterior \eqref{eq:ExpPosterior} with the prior \eqref{eq:ExpPrior} we note that the posterior indeed is of the same form as the prior, but with updated hyperparameters $\chi_{\text{new}}=\chi+s(z)$ and $\nu_{\text{new}}=\nu+1$. Hence, we have conjugacy for distributions in the exponential family, and the parameter posterior is given by $p(\eta|z,\chi,\nu)=\pi(\eta|\chi+s(z),\nu+1)$. We can obtain the likelihood of the data $z$ by marginalizing out the natural parameters $\eta$ from the joint distribution $p(z,\eta|\chi,\nu)=p(z|\eta)p(\eta|\chi,\nu)$ which gives 
\begin{align}
	p(z|\chi,\nu)&=\int p(z,\eta|\chi,\nu)\mathrm{d}\eta= h(z)g(\chi,\nu) \int \underbrace{  \exp \left( \eta^\T \left(s(z)+\chi\right) - a(\eta)\left(1+\nu\right)\right)}_{\text{Unnormalized $\pi(\eta|\chi_{\text{new}},\nu_{\text{new}})$ }} \mathrm{d}\eta\\
	&= h(z) \frac{g(\chi,\nu)}{g(\chi_{\text{new}},\nu_{\text{new}})}
\end{align}
where \eqref{eq:ExpPrior} and \eqref{eq:ExpLhood} were inserted in the second equality. Hence, for members of the exponential family there is a closed form expression for the distribution of the data when the parameters have been marginalized out.

\section{\normalsize Restricted exponential family} \label{sec:RestExpfam}
When working with SSMs we can run into problems when using the standard formulation for likelihoods and priors for the exponential family presented in Supplementary \ref{sec:Expfam}. The reason for this is that we typically have a dependence on the previous state $x_\sind{t-1}$ in the transition density that can result in a log-partition function which depends on this previous state. Put in the same general notation as in the previous section we get the likelihood 
\begin{equation}
	p(z | \zeta, \eta) = h(z,\zeta) \exp \left( \eta ^\T s(z,\zeta) - a(\eta, \zeta) \right)
\end{equation}
where the variable $\zeta$ is a known extra parameter (e.g the previous state $x_\sind{t-1}$) and we note that the log-partition function $a$ depends on this parameter. If we wish to formulate an exponential family prior for this likelihood we get 
\begin{equation}
	\pi(\eta | \chi, \nu, \zeta) = g(\chi, \nu) \exp \left( \eta ^\T \chi - a(\eta, \zeta)\nu \right),
\end{equation}
that is, we get a different prior depending on the value on $\zeta$ and, consequently, we cannot easily formulate a general prior distribution which is conjugate to the complete data likelihood. Instead we propose to use a restricted exponential family where the log-partition function is assumed to be separable into one parameter-dependent part and one $\zeta$-dependent part. The likelihood is given by
\begin{equation}
	p(z | \zeta, \eta) = h(z,\zeta) \exp \left( \eta ^\T s(z,\zeta) - A^\T(\eta)r(\zeta)  \right)
\end{equation}
where $A(\eta)$ is the restricted log-partition function and $r(\zeta)$ is some function which only depends on $\zeta$. A conjugate prior for this likelihood is  
\begin{equation}
	\pi(\eta | \chi, \nu, \zeta) = g(\chi, \nu) \exp \left( \eta ^\T \chi -  A^\T(\eta)\nu \right).
\end{equation}
The parameter posterior is 
\begin{equation}
	\pi(\eta|z,\zeta,\chi,\nu)\propto p(z|\zeta,\eta)p(\eta|\chi,\nu,\zeta) = h(z, \zeta)g(\chi,\nu)\exp\left( \eta^\T \left(\chi+s(z, \zeta)\right)-A^\T(\eta)(\nu+r(\zeta)) \right).
\end{equation}
We note that the posterior is of the same form as the prior but with updated hyperparameters $\chi_{\text{new}}=\chi+s(z,\zeta)$ and $\nu_{\text{new}}=\nu+r(\zeta)$. Comparing with the standard exponential family we note that the only difference is that the statistic now depends also on $\zeta$ and that we get a statistic, $r(\zeta)$, to update also for $\nu$.

\section{\normalsize Derivation of ancestor weights for the restricted exponential family} \label{sec:ASderiv}
For SSMs with joint state and observation likelihood in the restricted exponential family the likelihood is given by \eqref{eq:modelExp} and the parameter prior is $\pi(\theta|\chi_\sind0,\nu_\sind0) = g(\chi_\sind0,\nu_\sind0) \exp \left( \theta^\T \chi_\sind0 - A^\T(\theta)\nu_\sind0 \right)$. The ancestor weights are given by equation \eqref{eq:wPGASm}, which can be expanded to
\begin{equation} \label{eq:wAS}
	\begin{split}
		\tilde{w}^\sind i_\sind{t-1|T} \propto& \bar{w}^\sind i_\sind{t-1} \frac{p(x'_\sind{t:T},y_\sind{t:T}|x^\sind i_\sind{0:t-1},y_\sind{1:t-1})p(x^\sind i_\sind{0:t-1},y_\sind{1:t-1})}{p(x^\sind i_\sind{0:t-1},y_\sind{1:t-1})}=\bar{w}^\sind i_\sind{t-1} p(x'_\sind{t:T},y_\sind{t:T}|x^\sind i_\sind{0:t-1},y_\sind{1:t-1}) \\
		=& \bar{w}^\sind i_\sind{t-1} p(x'_\sind{t},y_\sind{t}|x^\sind i_\sind{0:t-1},y_\sind{1:t-1})\prod_{k=t+1}^{T} p(x'_\sind{k},y_\sind{k}|x^\sind i_\sind{0:t-1},y_\sind{1:t-1}, x'_\sind{t:k-1},y_\sind{t:k-1}) \\
		=& \bar{w}^\sind i_\sind{t-1} \int p(x'_\sind{t},y_\sind{t}|x^\sind i_\sind{t-1},\theta)p(\theta|x^\sind i_\sind{0:t-1},y_\sind{1:t-1})\mathrm{d}\theta \\
		&\hspace{1mm} \prod_{k=t+1}^T\int p(x'_\sind{k},y_\sind{k}|x'_\sind{k-1},\theta)p(\theta|x^\sind i_\sind{0:t-1},x'_\sind{t:k-1},y_\sind{1:k-1})\mathrm{d}\theta.
	\end{split}
\end{equation}
Now, to continue we need to compute the two integrals in the last equality for members in the restricted exponential family. First, note that
\begin{align} \label{eq:lhoodAnc}
	&p(x^\sind i_\sind{0:t-1},y_\sind{1:t-1}|\theta) = \big(\prod_{j=1}^{t-1} h_\sind j ^\sind i \big) \exp \big( \theta^\T \sum_{j=1}^{t-1}s_\sind j ^\sind i - A^\T(\theta) \sum_{j=1}^{t-1}r_\sind j ^\sind i\big)
\end{align}
and therefore
\begin{align} \label{eq:postAnc}
	p(\theta|x^\sind i_\sind{0:t-1},y_\sind{1:t-1}) &\propto p(x_\sind{0:t-1}^\sind i, y_\sind{1:t-1}|\theta)p(\theta|\chi,\nu) \\
	&= g(\chi_\sind0,\nu_\sind0)\big(\prod_{j=1}^{t-1} h_\sind j ^\sind i \big) \exp \Big( \theta^\T \big(\chi_\sind0 + \sum_{j=1}^{t-1}s_\sind j ^\sind i \big) - A^\T(\theta) \big( \nu_\sind0 + \sum_{j=1}^{t-1}r_\sind j ^\sind i\big) \Big) \\
	&\propto \pi(\theta|\chi_\sind{t-1}^\sind i,\nu_\sind{t-1}^\sind i),
\end{align}
where $\chi_\sind{t-1}^\sind i=\chi_\sind0 + \sum_{j=1}^{t-1}s_\sind j ^\sind i=\chi_\sind0 + s_\sind{1:t-1}^\sind i$ and $\nu_\sind{t-1}^\sind i=\nu_\sind0 + \sum_{j=1}^{t-1}r_\sind j ^\sind i=\nu_\sind0 + r_\sind{1:t-1}^\sind i$ and the last proportionality comes from comparison with the exponential factor in the prior. Similarly, by splitting in three terms (one for the ancestral part, one for the cross-over and one for the reference trajectory part) we have
\begin{align} \label{eq:lhoodRef}
&p(x^\sind i_\sind{0:t-1},x'_\sind{t:k-1},y_\sind{1:k-1}|\theta) = \Big(\prod_{j=1}^{t-1}p(x_\sind j^\sind i, y_\sind j |x_\sind{j-1}^\sind i,\theta )\Big) p(x_\sind t', y_\sind t|x_\sind{t-1}^\sind i,\theta ) \Big(\prod_{j=t+1}^{k-1}p(x_\sind j', y_\sind j|x_\sind{j-1}',\theta ) \Big) \\
&= \underbrace{\big(\prod_{j=1}^{t-1} h_\sind j ^\sind i \big) h'_\sind t \big(\prod_{j=t+1}^{k-1} h'_\sind j \big)}_{H} \exp \Big( \theta^\T \big( \underbrace{s_\sind{1:t-1}^\sind i + s_\sind t (x'_\sind t, x_\sind{t-1}^\sind i, y_\sind t) + s'_\sind{t+1:k-1}}_{ S }  \big) \\
& \hspace{2mm} -A^\T(\theta) \big( \underbrace{r_\sind{1:t-1}^\sind i + r_\sind t (x_\sind{t-1}^\sind i) + r'_\sind{t+1:k-1}}_{ R } \big) \Big)
\end{align}
and therefore, using the abbreviations $H$,$S$, and $R$ above, the parameter posterior is
\begin{align} \label{eq:postRef}
p(\theta|x^\sind i_\sind{0:t-1},x'_\sind{t:k-1},y_\sind{1:k-1}) &\propto g(\chi_\sind0,\nu_\sind0) H  \exp \Big( \theta^\T \big(\chi_\sind0 + S  \big) - A^\T(\theta) \big( \nu_\sind0 + R \big) \Big) \\
& \propto \pi(\theta|\chi_\sind{k-1}^\sind i,\nu_\sind{k-1}^\sind i)
\end{align}
where
\begin{equation}
	\begin{split}
		&\chi_\sind{k-1}^\sind i=\chi_\sind0 + s_\sind{1:t-1}^\sind i + s_\sind t (x'_\sind t, x_\sind{t-1}^\sind i, y_\sind t) + s'_\sind{t+1:k-1} = \chi_\sind{t-1}^\sind i + s_\sind t (x'_\sind t, x_\sind{t-1}^\sind i, y_\sind t) + s'_\sind{t+1:k-1} \\
		&\nu_\sind{k-1}^\sind i=\nu_\sind0 + r_\sind{1:t-1}^\sind i + r_\sind t (x_\sind{t-1}^\sind i) + r'_\sind{t+1:k-1}=\nu_\sind{t-1}^\sind i + r_\sind t (x_\sind{t-1}^\sind i) + r'_\sind{t+1:k-1}.
	\end{split}
\end{equation}
We can now compute the integrals in \eqref{eq:wAS}:
\begin{equation} \label{eq:Int1}
	\begin{split}
		&\int p(x'_\sind{t},y_\sind{t}|x^\sind i_\sind{t-1},\theta)p(\theta|x^\sind i_\sind{0:t-1},y_\sind{1:t-1})\mathrm{d}\theta = \\
		& \int h_\sind t\exp \big( \theta^\T s'_\sind t - A^\T(\theta)r'_\sind t \big) g(\chi_\sind{t-1} ^\sind i,\nu_\sind{t-1} ^\sind i ) \exp \big( \theta^\T \chi_\sind{t-1}^\sind i - A^\T(\theta)\nu_\sind{t-1}^\sind i \big) \mathrm{d}\theta \\
		&= h_\sind t g(\chi_\sind{t-1} ^\sind i,\nu_\sind{t-1} ^\sind i ) \int \underbrace{ \exp \big( \theta^\T(s'_\sind t + \chi_\sind{t-1}^\sind i ) - A^\T(\theta)(r'_\sind t + \nu_\sind{t-1}^\sind i) \big) }_{\text{Unnormalized $\pi(\theta|\chi_\sind t ^\sind i ,\nu_\sind t^\sind i)$}} \mathrm{d}\theta \\ 
		&= h_\sind t \frac{g(\chi_\sind{t-1} ^\sind i,\nu_\sind{t-1} ^\sind i )}{g(\chi_\sind t ^\sind i,\nu_\sind t^\sind i)}.
	\end{split}
\end{equation}
Note that $h_\sind t = h_\sind t (x'_\sind t, x_ \sind{t-1}^i, y_\sind t)$ depends on the ancestral path. In a similar fashion, for the second integral we obtain
\begin{equation} \label{eq:Int2}
	\begin{split}
		&\int p(x'_\sind{k},y_\sind{k}|x'_\sind{k-1},\theta)p(\theta|x^\sind i_\sind{0:t-1},x'_\sind{t:k-1},y_\sind{1:k-1})\mathrm{d}\theta = \\
		& h'_\sind k g(\chi_\sind{k-1} ^\sind i,\nu_\sind{k-1} ^\sind i ) \int \exp \big( \theta^\T(s'_\sind k + \chi_\sind{k-1}^\sind i ) - A^\T(\theta)(r'_\sind k + \nu_\sind{k-1}^\sind i) \big) \mathrm{d}\theta \\
		&= h'_\sind k \frac{g(\chi_\sind{k-1} ^\sind i,\nu_\sind{k-1} ^\sind i )}{g(\chi_\sind k ^\sind i,\nu_\sind k^\sind i)},
	\end{split}
\end{equation}
where $h'_\sind k = h'_\sind k (x'_\sind k, x'_ \sind{k-1}, y_\sind t)$ only depends on the reference trajectory and observations. Now, by substituting \eqref{eq:Int1} and \eqref{eq:Int2} into \eqref{eq:wAS}, we obtain
\begin{equation}
	\begin{split}
		\tilde{w}^\sind i_\sind{t-1|T} \propto& \bar{w}^\sind i_\sind{t-1}h_\sind t \frac{g(\chi_\sind{t-1} ^\sind i,\nu_\sind{t-1} ^\sind i )}{g(\chi_\sind t ^\sind i,\nu_\sind t^\sind i)} \prod_{k=t+1}^{T}h'_\sind k \frac{g(\chi_\sind{k-1} ^\sind i,\nu_\sind{k-1} ^\sind i )}{g(\chi_\sind k ^\sind i,\nu_\sind k^\sind i)} \\
		&\propto \bar{w}^\sind i_\sind{t-1}h_\sind t \frac{g(\chi_\sind{t-1} ^\sind i,\nu_\sind{t-1} ^\sind i )}{g(\chi_\sind T ^\sind i,\nu_\sind T^\sind i)},
	\end{split}
\end{equation}
where the (surprisingly) simple final expression is due to all terms except $g(\chi_\sind{t-1} ^\sind i,\nu_\sind{t-1} ^\sind i )$ and $g(\chi_\sind T ^\sind i,\nu_\sind T^\sind i)$ canceling in the product. Note also that $h'_\sind k$ can be removed (in proportionality) since they are the same for all ancestor particles.

\section{\normalsize Theoretical justification of the blocking scheme} \label{sec:JustBlocking}
To show that the blocking scheme in Algorithm \ref{alg:Blocking} is correct we start by verifying that the underlying (hypothetical) partially collapsed Gibbs (PCG) sampler is correct. Following the notation in \cite{vanDyk08} we set $X_\sind 1= x_\sind{0:B}$, $X_\sind 2 = x_\sind{B+1:B+L}$ (the overlap), $X_\sind 3=x_\sind{B+L+1:T}$, $Y=y_\sind{1:T}$, and the superscript $^*$ indicates intermediate quantities. The joint distribution we wish to sample from is $p\left(X_\sind 1, X_\sind 2, X_\sind 3, \theta|Y\right)$. The underlying hypothetical PCG of the blocking scheme is
\begin{equation} \label{eq:IdealBlocked}
	\begin{split}
		X_\sind 1, X^*_\sind 2 & \sim p\left( X_\sind 1, X_\sind 2| X_\sind 3, Y, \theta \right) \\
		X_\sind 2, X_\sind 3 & \sim p\left( X_\sind 2, X_\sind 3 | X_\sind 1, Y \right)  \\
		\theta & \sim p\left( \theta | X_\sind 1, X_\sind 2, X_\sind 3, Y\right).
	\end{split}
\end{equation}
We note that step 1 and 3 are correct hypothetical Gibbs steps, the only issue is the marginalized step 2. However, in step 2 it is possible to also sample $\theta$ from its full conditional, that is to sample $X_\sind 2, X_\sind 3,\theta^*  \sim p\left( X_\sind 2, X_\sind 3 | X_\sind 1, Y \right)p\left(\theta|X_\sind 1, X_\sind 2, X_\sind 3, Y \right)=p\left( X_\sind 2, X_\sind 3, \theta | X_\sind 1, Y \right)$. This sampling of $\theta^*$ is redundant since the sampled value is never conditioned on in the following step and can be removed according to the reasoning in \cite{vanDyk08}, concluding that the underlying hypothetical PCG of the blocking scheme is indeed a correct hypothetical PCG. If one or more of the conditionals in \eqref{eq:IdealBlocked} are not possible to sample from directly some MCMC scheme (such as PG/mPG) can be used. Describing the PMCMC steps in terms of kernels, $K$, Algorithm \ref{alg:Blocking} can be formulated
\begin{equation} \label{eq:MCMCBlocked}
	\begin{split}
		X_\sind 1, X^*_\sind 2 & \sim K\left( X_\sind 1, X_\sind 2| X'_\sind 1, X'_\sind 2; X_\sind 3, Y, \theta \right) \\
		X_\sind 2, X_\sind 3 & \sim K\left( X_\sind 2, X_\sind 3 | X^*_\sind 2, X'_\sind 3; X_\sind 1, Y \right)  \\
		\theta & \sim p\left( \theta | X_\sind 1, X_\sind 2, X_\sind 3, Y\right)
	\end{split}
\end{equation}
where $'$ indicates the (reference) trajectory from the previous iteration. Again, step 1 and 3 are correct (as long as $K$ in step 1 is correct, e.g. a PG kernel), the main concern is step 2. However, as was the case for the hypothetical sampler, it is possible to add the sampling of $\theta$ in step 2. Sampling $\theta$ is, again, redundant and step 2 is also correct provided that the stationary distribution of $K$ is the marginalized target.

\section{\normalsize Additional results} \label{sec:AddResult}
In this section we provide some additional results from numerical simulations. 

\subsection{Toy model in Section \ref{sec:mPG}}

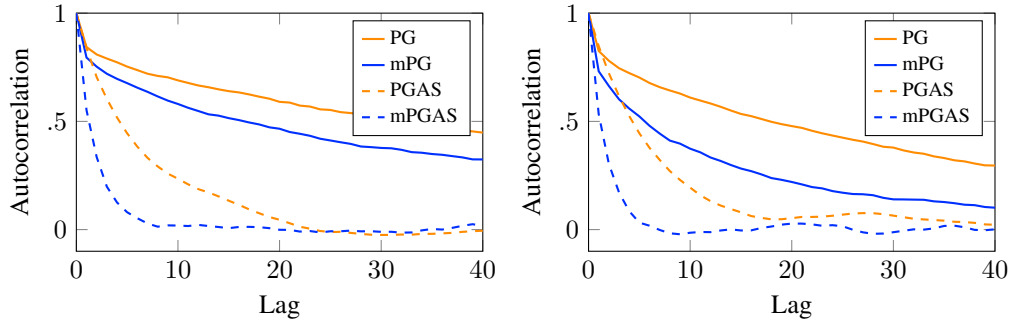
\begin{figure}[ht]
	\begin{center}
		\begin{tikzpicture}[scale=1, transform shape]
  \pgfplotstableread[col sep=comma]{data/acfdata_matlab.csv}{\dataM}
  \pgfplotstableread[col sep=comma]{data/autocorr_toy_model.csv}{\acfdata}
  \begin{axis}[
  xmin=0,
  xmax=40,
  ymin=-0.1,
  ymax=1,
  xlabel=Lag,
  ylabel=Autocorrelation,
  width = 0.5\textwidth,
  legend columns = 1,
  legend pos = north east,
  name = myplot
  ]
  \addplot[thick, yellow1] table[x=lag, y=PG] {\dataM};
  \addplot[thick, blue1] table[x=lag, y=mPG] {\dataM};
  \addplot[thick, dashed, yellow1] table[x=lag, y=PGAS] {\dataM};
  \addplot[thick, dashed, blue1] table[x=lag, y=mPGAS] {\dataM};
  \legend{PG, mPG, PGAS, mPGAS};
  \end{axis}
  \begin{axis}[
  xmin=0,
  xmax=40,
  ymin=-0.1,
  ymax=1,
  xlabel=Lag,
  ylabel=Autocorrelation,
  width = 0.5\textwidth,
  legend columns = 1,
  legend pos = north east,
  at = (myplot.north east),
  anchor=north west,
  xshift=4em,
  ]
  \addplot[thick, yellow1] table[x=lag, y=s2x_PG] {\acfdata};
  \addplot[thick,blue1] table[x=lag, y=s2x_mPG] {\acfdata};
  \addplot[thick,  dashed, yellow1] table[x=lag, y=s2x_PGAS] {\acfdata};
  \addplot[thick, dashed, blue1] table[x=lag, y=s2x_mPGAS] {\acfdata};
  \legend{PG, mPG, PGAS, mPGAS};
  \end{axis}
\end{tikzpicture}
	\end{center}
	\caption{Results of the simulation of the model \eqref{eq:toyModel}. \emph{Left:} autocorrelation function for all methods, obtained from simulations in Matlab. \emph{Right:} autocorrelation function for all methods, obtained from simulations in Birch. }
	\label{fig:toyBirchMatlab}
\end{figure}

Figure \ref{fig:toyBirchMatlab} shows the results for our implementation of \eqref{eq:toyModel} in Matlab (left) and Birch (right). We generated $T=250$ observations from \eqref{eq:toyModel} with $\sigma_\sind{v}^{2}=5.3$ and $\sigma_\sind{w}^{2}=9$. We used hyperparameters $\alpha_\sind v = \alpha_\sind w = 2$ and $ \beta_\sind v = \beta_\sind w =10$, and all four methods were run for $M=10000$ iterations. We initialized with $\sigma_\sind v^{2}=4.3$ and $\sigma_\sind w^2=9.4$ and used a bootstrap proposal for PG/PGAS and a marginalized bootstrap proposal for mPG/mPGAS. We note that for both implementations there is a clear improvement from marginalizing out the parameters (for both PG and PGAS), and that it is beneficial to use PGAS/mPGAS rather than PG/mPG for this model.

\subsection{Vector-borne disease model in Section \ref{sec:PPL}}
\begin{figure}[t]
	\begin{center}
		\begin{tikzpicture}[scale=1, transform shape]
  \pgfplotstableread[col sep=comma]{data/density.csv}{\datavbddensity}
  \pgfplotstableread[col sep=comma]{data/histogram.csv}{\datavbdhist}
  \pgfplotstableread[col sep=comma]{data/autocorr.csv}{\datavbdautocorr}
  \begin{axis}[
  xmin=0.1,
  xmax=0.8,
  xlabel=Reporting rate,
  width = 0.5\textwidth,
  ylabel=Density,
  legend columns = 1,
  legend pos = north east,
  name = myplot
  ]
  \addplot[ybar interval, fill=yellow1!50, area legend] table[x=rho, y=density_pgas] {\datavbdhist};
  \addplot[thick, blue1] table[x=rho,y=density_mpgas] {\datavbddensity};
  \addplot[very thick, densely dashed, black] table[x=rho, y=density_imp]{\datavbddensity};
  \legend{PGAS, mPGAS, mIS};
  \end{axis}
  \begin{axis}[
  	xmin=0,
  	xmax=80,
  	ymin=-0.1,
  	ymax=1,
  	xlabel=Lag,
  	width = 0.5\textwidth,
  	ylabel=Autocorrelation,
	legend columns = 1,
	legend pos = south west,
	at = (myplot.north east),
	anchor=north west,
	xshift=4em,
  ]
  \addplot[thick,  yellow1] table[x=lag, y=pg0] {\datavbdautocorr};
  \addplot[thick,  blue1] table[x=lag, y=mpg0] {\datavbdautocorr};
  \addplot[thick, densely dashed, yellow1] table[x=lag, y=pgas0] {\datavbdautocorr};
  \addplot[thick, densely dashed,blue1] table[x=lag, y=mpgas0] {\datavbdautocorr};
  \addplot[thick,  yellow1] table[x=lag, y=pg1] {\datavbdautocorr};
  \addplot[thick,  yellow1] table[x=lag, y=pg2] {\datavbdautocorr};
  \addplot[thick,  yellow1] table[x=lag, y=pg3] {\datavbdautocorr};
  \addplot[thick,  blue1] table[x=lag, y=mpg1] {\datavbdautocorr};
  \addplot[thick,  blue1] table[x=lag, y=mpg2] {\datavbdautocorr};
  \addplot[thick,  blue1] table[x=lag, y=mpg3] {\datavbdautocorr};
  \addplot[thick, densely dashed,yellow1] table[x=lag, y=pgas1] {\datavbdautocorr};
  \addplot[thick, densely dashed,yellow1] table[x=lag, y=pgas2] {\datavbdautocorr};
  \addplot[thick, densely dashed,yellow1] table[x=lag, y=pgas3] {\datavbdautocorr};
  \addplot[thick, densely dashed,blue1] table[x=lag, y=mpgas1] {\datavbdautocorr};
  \addplot[thick, densely dashed,blue1] table[x=lag, y=mpgas2] {\datavbdautocorr};
  \addplot[thick, densely dashed,blue1] table[x=lag, y=mpgas3] {\datavbdautocorr};
  \legend{PG, mPG, PGAS, mPGAS};
  \end{axis}
\end{tikzpicture}
	\end{center}
	\caption{Results of the simulation of the vector-borne disease model. \emph{Left:} estimated density of the reporting rate parameter for PGAS and mPGAS, mean of four chains. For comparison, a run of marginalized importance sampling (mIS) is also shown. \emph{Right:} estimated autocorrelation function of the reporting rate parameter for all methods, four chains for each method.}
	\label{fig:VBD-pgas}
\end{figure}
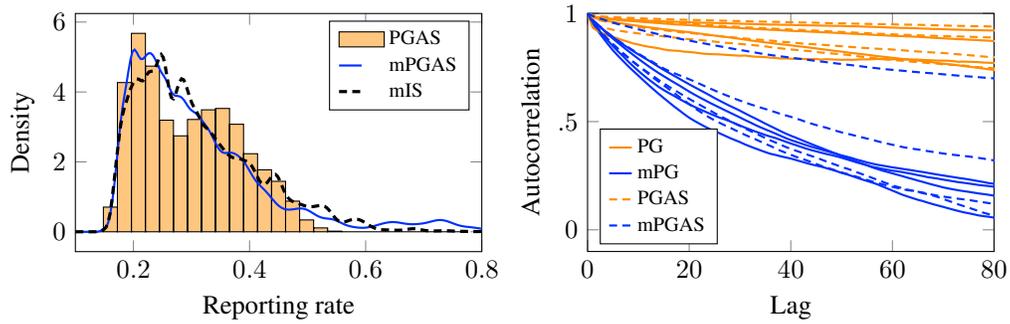

Figure \ref{fig:VBD-pgas} shows the results of a simulation of the vector-borne disease model in Birch of four different chains; for each of these 10000 samples were drawn using PGAS and mPGAS. The samplers used $N=1024$ particles each. The autocorrelation of the samples is improved by marginalizing out the parameters. However, we note that there is no obvious improvement from using PGAS rather than PG for this model. One explanation for this could be that it is difficult to match the reference trajectory with an ancestor which is not the reference trajectory for this compartmental model. The reference trajectory would then be sampled in the ancestor sampling step, which in turn means that PGAS would reduce to PG.

\end{document}